\documentclass[12pt]{article}
\usepackage{a4wide,epsfig,psfrag,amsmath,amssymb,cite,scalefnt}
\usepackage[dvipsnames]{xcolor}
\usepackage{amsmath,comment,braket}
\usepackage{placeins}
\usepackage{breqn}
\usepackage{slashed}
\usepackage{comment}
\usepackage{subcaption}

\graphicspath{ {figs/} }

\allowdisplaybreaks

\begin{document}

\title{\vskip-3cm{\baselineskip14pt
    \begin{flushleft}
      \normalsize P3H-20-038, TTP20-028
    \end{flushleft}} \vskip1.5cm 
  Exact results for $Z_m^{\rm OS}$ and $Z_2^{\rm OS}$ with two mass scales and
  up to three loops}

\author{
  Matteo Fael, Kay Sch\"onwald and Matthias Steinhauser
  \\[1em]
  {\small\it Institut f{\"u}r Theoretische Teilchenphysik,
    Karlsruhe Institute of Technology (KIT)}\\
  {\small\it 76128 Karlsruhe, Germany}  
}
  
\date{}

\maketitle

\thispagestyle{empty}

\begin{abstract}

  We consider the on-shell mass and wave function renormalization constants
  $Z_m^{\rm OS}$ and $Z_2^{\rm OS}$ up to three-loop order allowing for a
  second non-zero quark mass.  We obtain analytic results in terms of harmonic
  polylogarithms and iterated integrals with the additional letters
  $\sqrt{1-\tau^2}$ and $\sqrt{1-\tau^2}/\tau$ which extends the findings from
  Ref.~\cite{Bekavac:2007tk} where only numerical expressions are presented.
  Furthermore, we provide terms of order ${\cal O}(\epsilon^2)$ and
  ${\cal O}(\epsilon)$ at two- and three-loop order which are crucial
  ingrediants for a future four-loop calculation. Compact results for the
  expansions around the zero-mass, equal-mass and large-mass cases allow for a
  fast high-precision numerical evaluation.

\end{abstract}


\thispagestyle{empty}

\newpage


\section{Introduction and notation}

Once quantum corrections to quantities, which involve heavy quarks, are
computed to higher orders in perturbation theory the renormalization of the
mass and wave function has to be performed. The corresponding renormalization
constants are usually denoted by $Z_m^{\rm OS}$ and $Z_2^{\rm OS}$,
respectively. They are defined through
\begin{eqnarray}
  m^0 &=& Z_m^{\rm OS} m^{\rm OS}\,,\nonumber\\
  \psi^0 &=& \sqrt{Z_2^{\rm OS}} \psi^{\rm OS}\,,
\end{eqnarray}
where $m^0$ and $\psi^0$ stand for the bare quark mass and wave function.
The superscript ``OS'' refers to the on-shell scheme, which for QCD
corrections is used synonymous to the pole scheme.

\begin{figure}[b]
  \centering
    \includegraphics[width=0.18\linewidth]{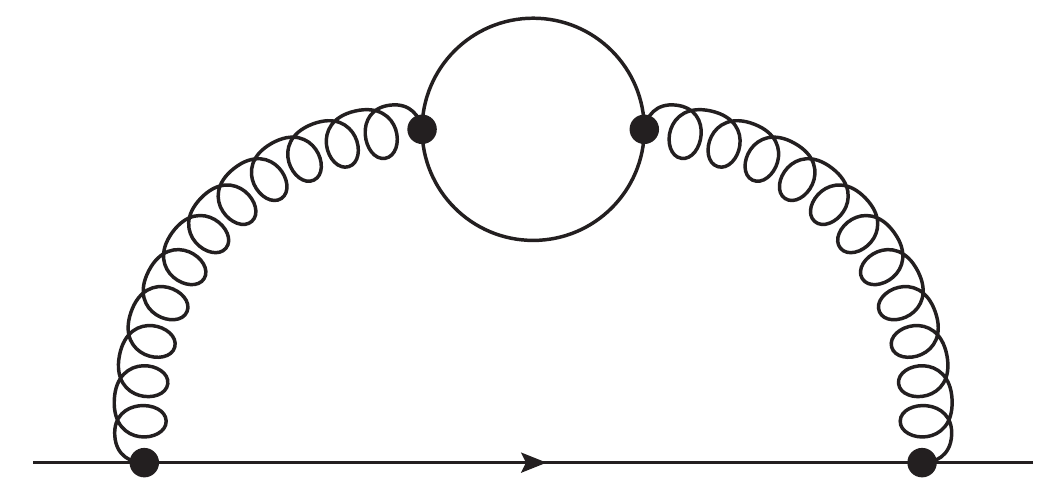}
    \includegraphics[width=0.18\linewidth]{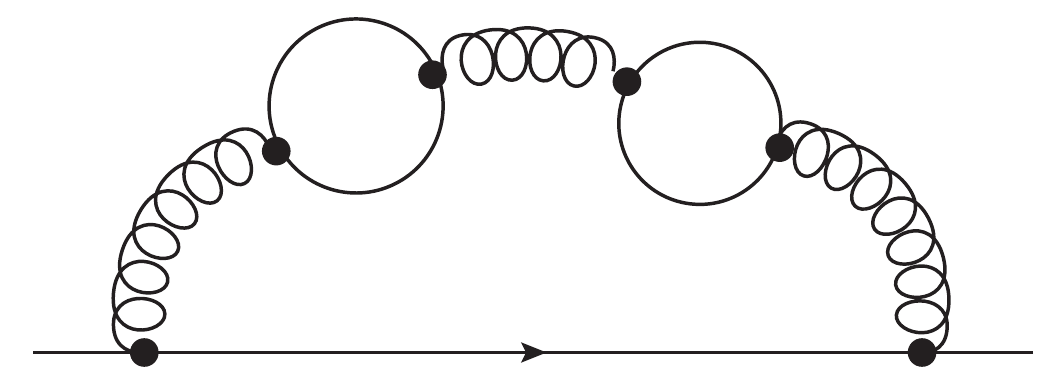}
    \includegraphics[width=0.18\linewidth]{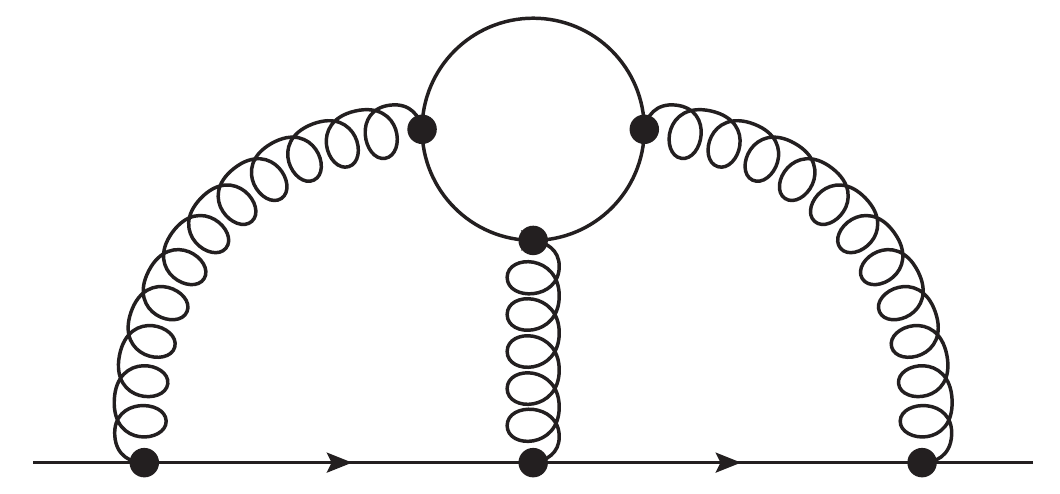}
    \includegraphics[width=0.18\linewidth]{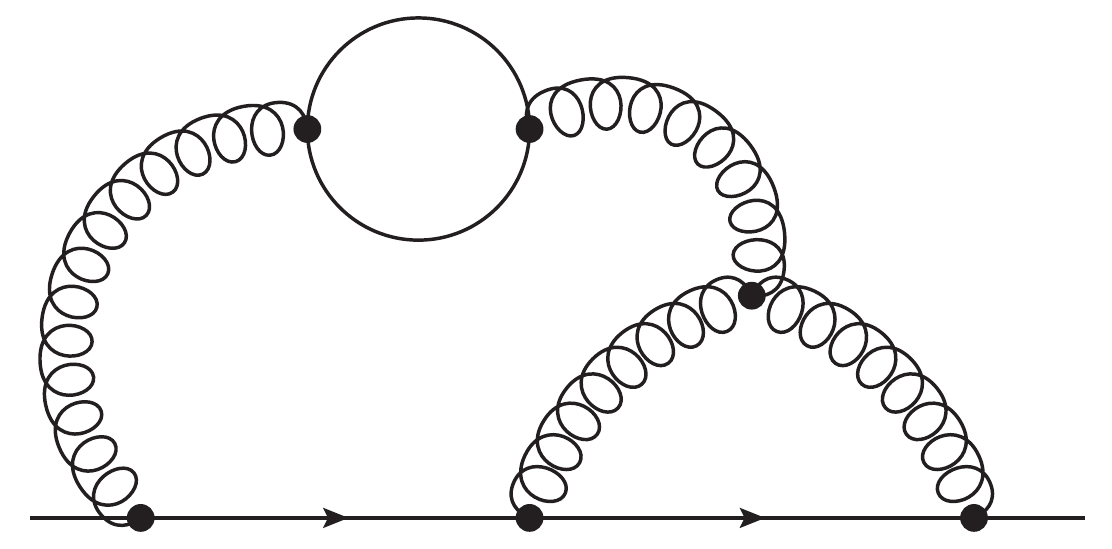}
    \includegraphics[width=0.18\linewidth]{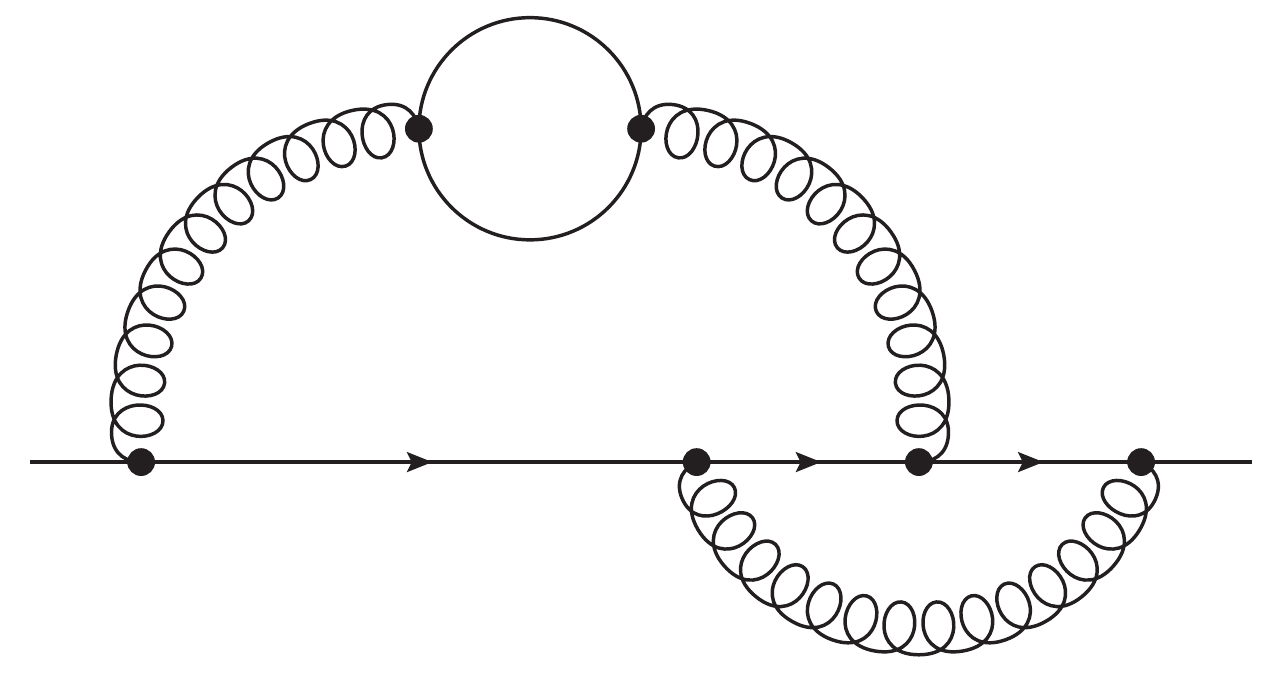}
  \caption{\label{fig::FDs}Sample Feynman diagrams contributing to 
  $Z_m^{\rm OS}$ and $Z_2^{\rm OS}$. Straight and curly lines represent
  quarks and gluons, respectively. The fermions in the closed loop
  may have a different mass from the external fermion.}
\end{figure}

Within QCD, analytic results for both renormalization constants are available
up to three
loops~\cite{Tarrach:1980up,Gray:1990yh,Broadhurst:1991fy,Chetyrkin:1999ys,Chetyrkin:1999qi,Melnikov:2000qh,Melnikov:2000zc,Marquard:2007uj}.
At four-loop
order~\cite{Marquard:2015qpa,Marquard:2016dcn,Marquard:2018rwx,Laporta:2020fog}
semi-analytic methods were used.  Starting from two loops there are
contributions with closed quark loops, which can either be massless, have the
mass of the external quark ($m_1$), or have a different mass ($m_2$). 
Sample Feynman diagrams of this type can be found in Fig.~\ref{fig::FDs}. The case
$0\not= m_2\not=m_1$ was considered in
Refs.~\cite{Gray:1990yh,Broadhurst:1991fy} and~\cite{Bekavac:2007tk} at two-
and three-loop orders.  In this work we re-consider these contributions to
$Z_m^{\rm OS}$ and $Z_2^{\rm OS}$ up to three-loop order and provide analytic
results including ${\cal O}(\epsilon)$ terms. In
Ref.~\cite{Bekavac:2007tk} only expansions for $m_2/m_1\to0$ and numerical
results have been provided up to the constant term in $\epsilon$.

It is convenient to introduce the variable
\begin{eqnarray}
  x & = & \frac{m_2}{m_1}\,.
\end{eqnarray}
We furthermore adopt the notation from~\cite{Bekavac:2007tk} and write
($i=m,2$)
\begin{eqnarray}
  Z_i^{\rm OS} &=& 1 + \frac{\alpha_s(\mu)}{\pi} \left(\frac{e^{\gamma_E}}{4 \pi}
  \right)^{-\epsilon} \delta Z_i^{(1)} +
  \left(\frac{\alpha_s(\mu)}{\pi}\right)^2 \left(\frac{e^{\gamma_E}}{4 \pi}
  \right)^{-2\epsilon} \delta Z_i^{(2)} \nonumber \\
  && + \left(\frac{\alpha_s(\mu)}{\pi}\right)^3
  \left(\frac{e^{\gamma_E}}{4\pi} \right)^{-3\epsilon} \delta Z_i^{(3)}
  + \mathcal{O}\left(\alpha_s^4\right) \,,
  \label{eq::mass::Zm}
\end{eqnarray}
where $\alpha_s$ denotes the $\overline{\rm MS}$ renormalized strong coupling constant defined in
$n_f$-flavour QCD, and $\mu$ is the renormalization constant; $\gamma_E$ is
the Euler-Mascheroni constant.  We decompose the coefficients
$\delta Z_i^{(k)}$ according to the different colour factors and obtain
\begin{eqnarray}
  \delta Z_i^{(1)} &=& C_F \, Z_i^F
  \nonumber\\
  \delta Z_i^{(2)} &=& C_F^2\, Z_i^{FF} + C_FC_A\, Z_i^{FA} +
  + C_FT_Fn_l Z_i^{FL} + C_FT_Fn_h Z_i^{FH} + C_FT_Fn_m Z_i^{FM}(x) 
  \nonumber\\
  \delta Z_i^{(3)} &=& C_F^3\, Z_i^{FFF} + C_F^2C_A\, Z_i^{FFA} +
  C_FC_A^2\, Z_i^{FAA} 
  + C_FT_Fn_l \left( C_F\, Z_i^{FFL} + C_A\, Z_i^{FAL}
  \right.
  \nonumber\\&&\mbox{} 
  \left.
  + T_Fn_l\,Z_i^{FLL} + T_Fn_h\, Z_i^{FHL} + T_Fn_m\, Z_i^{FML}(x)
  \right) 
  \nonumber\\&&\mbox{} 
  + C_FT_Fn_h \left( C_F\, Z_i^{FFH} + C_A\, Z_i^{FAH} 
  + T_Fn_h\,Z_i^{FHH} + T_Fn_m\,Z_i^{FMH}(x) \right) \nonumber\\
  &&\mbox{} + C_FT_Fn_m \left( C_F\, Z_i^{FFM}(x) + C_A\, Z_i^{FAM}(x) 
  + T_Fn_m\,Z_i^{FMM}(x) \right)
  \,,
  \label{eq::mass::Zm3l}
\end{eqnarray}
with the SU$(N_c)$ colour factors $C_F=(N_c^2-1)/(2N_c)$, $C_A=N_c$ and
$T_F=1/2$. We have introduced the quantities $n_l$, $n_h$ and $n_m$
to label closed quark loops with mass zero, $m_1$ and $m_2$, respectively.
We have $n_f=n_l+n_m+n_h=n_l+1+1$ active quark flavours.
Note that only the terms proportional to $n_m$ and $n_m^2$ have a non-trivial
dependence on $x$. This is the main subject of the present paper.

For the quark mass renormalization constant we also introduce the ratio
\begin{eqnarray}
  z_m &=& \frac{Z_m^{\rm OS}}{Z_m^{\overline{\rm MS}}}
\end{eqnarray}
which is finite since both $Z_m^{\rm OS}$ and the $\overline{\rm MS}$
renormalization constant $Z_m^{\overline{\rm MS}}$ only contain ultra-violet
poles, which cancel in the ratio.  Note that $Z_2^{\rm OS}$ contains both
ultra-violet and infra-red poles. $z_m$ has an analogue perturbative expansion
as $Z_m^{\rm OS}$ and $Z_2^{\rm OS}$ in Eq.~(\ref{eq::mass::Zm}).

In Ref.~\cite{Bekavac:2007tk} the $n_m$-dependent terms of
$Z_m^{\rm OS}$ and $Z_2^{\rm OS}$ were computed up to three loops.  At
two-loop order analytic results were obtained. However, at three-loop
order, for the complicated master integrals only an expansion for $x\to0$
could be obtained. For larger values of $x$ a numerical evaluation was
necessary. For most practical purposes this is sufficient. However, in some
cases analytic expressions or expansions are useful. In this work we extend
the result of~\cite{Bekavac:2007tk} in the following aspects:
\begin{itemize}
\item We extend the $\epsilon$ expansion by one order both at two and three
  loops, which is necessary input for a future four-loop calculation
  of the $n_m$ terms of the on-shell renormalization constants.
\item We provide analytic results in terms of iterated integrals with the
  letters $\tau, 1-\tau, 1+\tau, \sqrt{1-\tau^2}, \sqrt{1-\tau^2}/\tau$. They are
  present both in the $\alpha_s^3\epsilon^0$ and $\alpha_s^3\epsilon^1$ terms.
\item We provide 26 terms in an anlytic expansions both for $x\to0$,
  $x\to1$ and $x\to\infty$ (i.e. up to order $x^{25}$, $(1-x)^{25}$ and
  $1/x^{25}$). 
  In Ref.~\cite{Bekavac:2007tk}, for the three-loop
  term only the expansion for $x\to0$ up to $x^8$ was considered.
\end{itemize}

In the next section we briefly describe the approach which we use to obtain
the analytic results and the expansions in the various limits.
In Section~\ref{sec::res} we discuss our results for the
renormalization constants and give our conclusions. In the Appendix we
provide details to the ancillary file of our paper~\cite{progdata}.


\section{Technicalities}

\begin{figure}[t]
  \centering
  \begin{subfigure}{0.3\textwidth}
    \centering
    \includegraphics[width=0.6\linewidth]{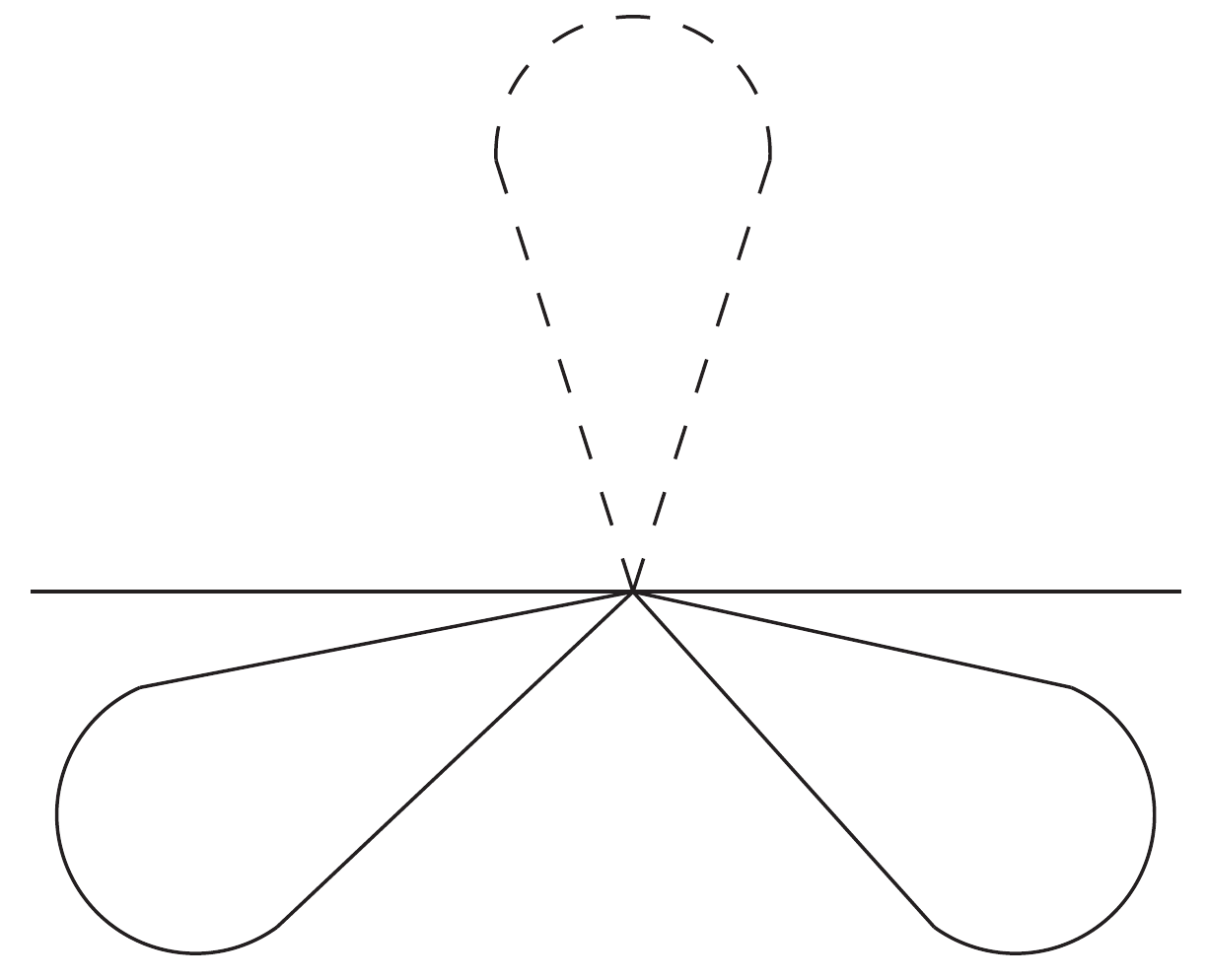}
    \caption*{$M_1$}
  \end{subfigure}
  \begin{subfigure}{0.3\textwidth}
    \centering
    \includegraphics[width=0.6\linewidth]{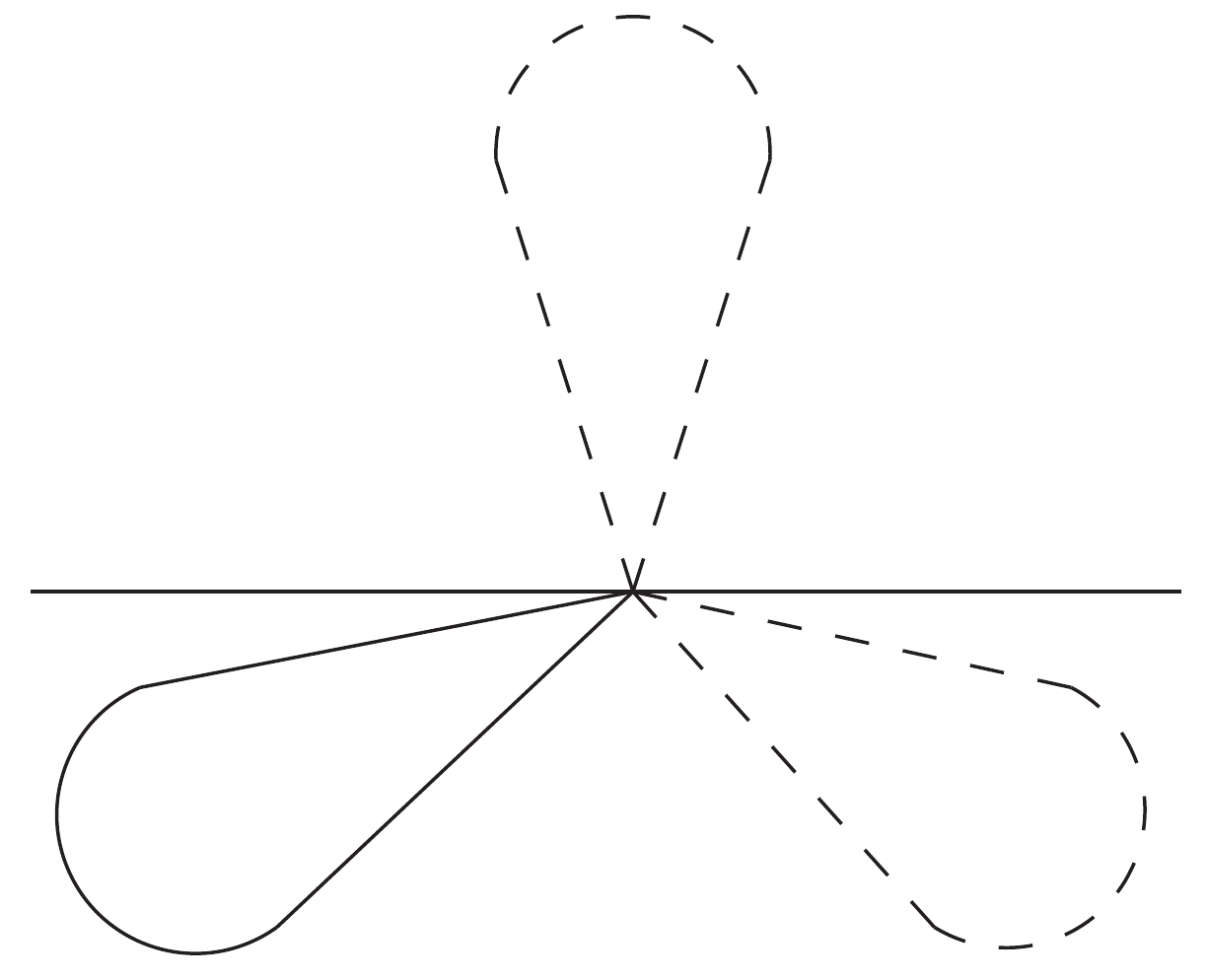}
    \caption*{$M_2$}
  \end{subfigure}
  \begin{subfigure}{0.3\textwidth}
    \centering
    \includegraphics[width=0.6\linewidth]{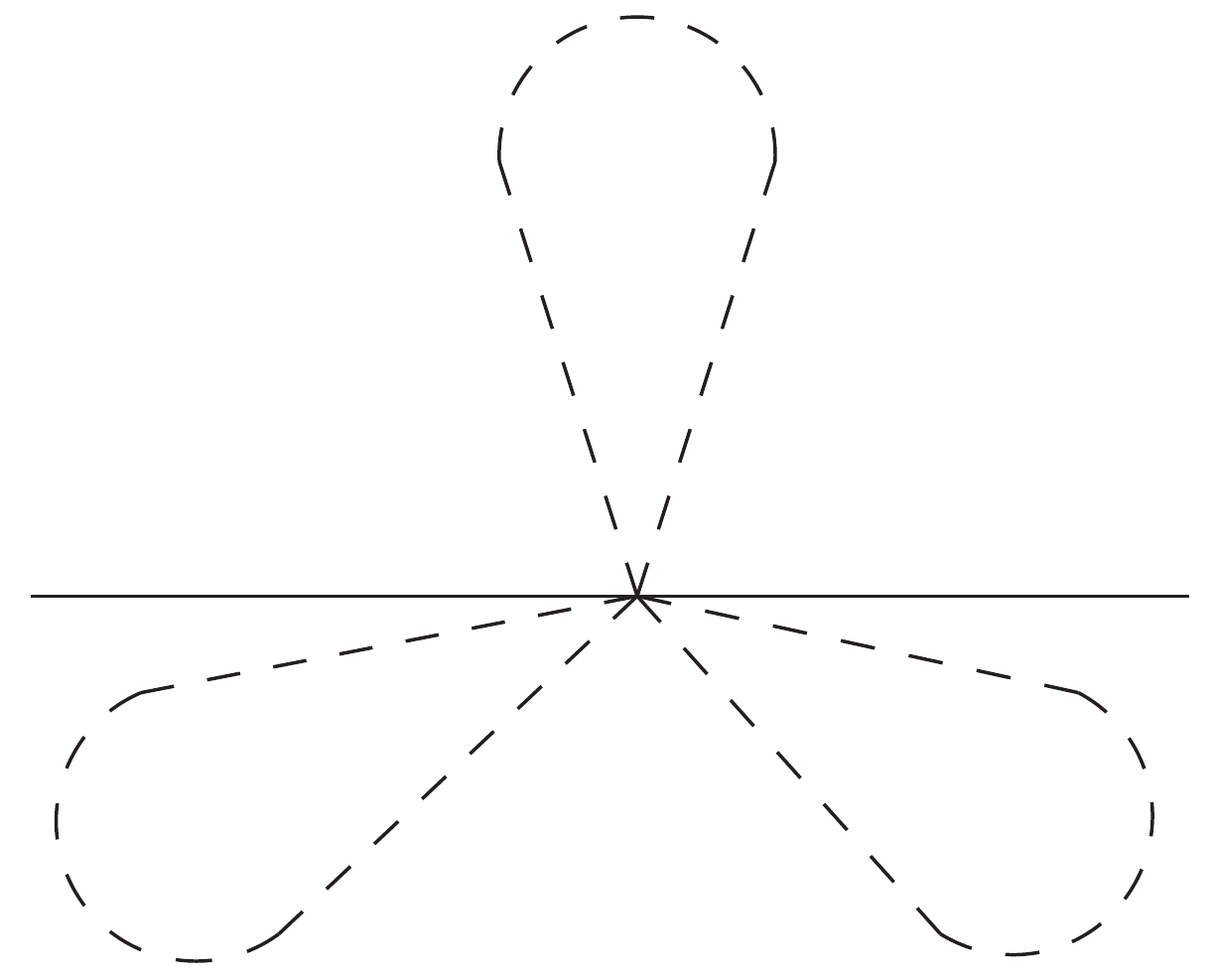}
    \caption*{$M_3$}
  \end{subfigure}
  \begin{subfigure}{0.23\textwidth}
    \centering
    \includegraphics[width=0.55\linewidth]{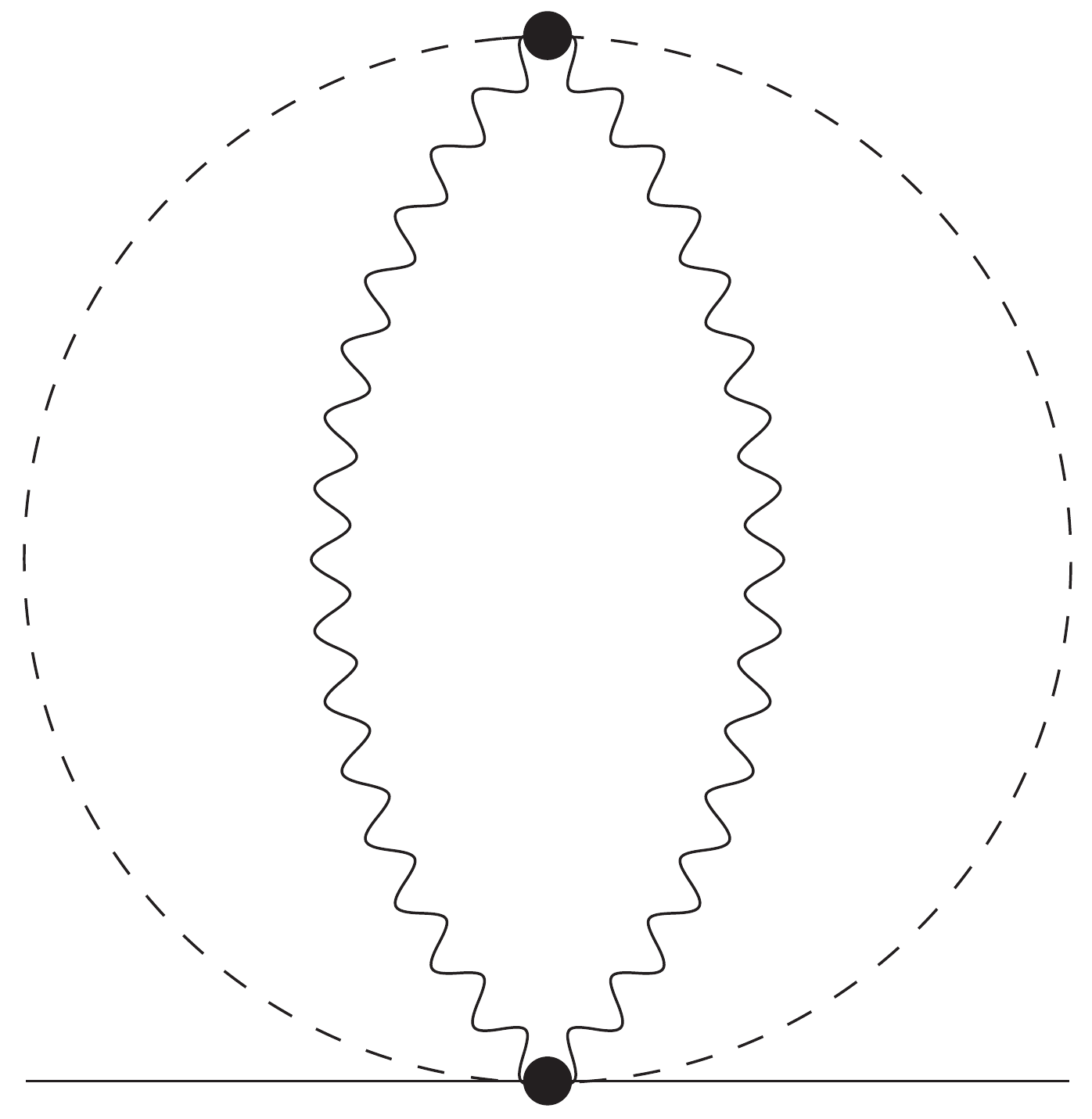}
    \caption*{$M_4$}
  \end{subfigure}
  \begin{subfigure}{0.23\textwidth}
    \centering
    \includegraphics[width=0.55\linewidth]{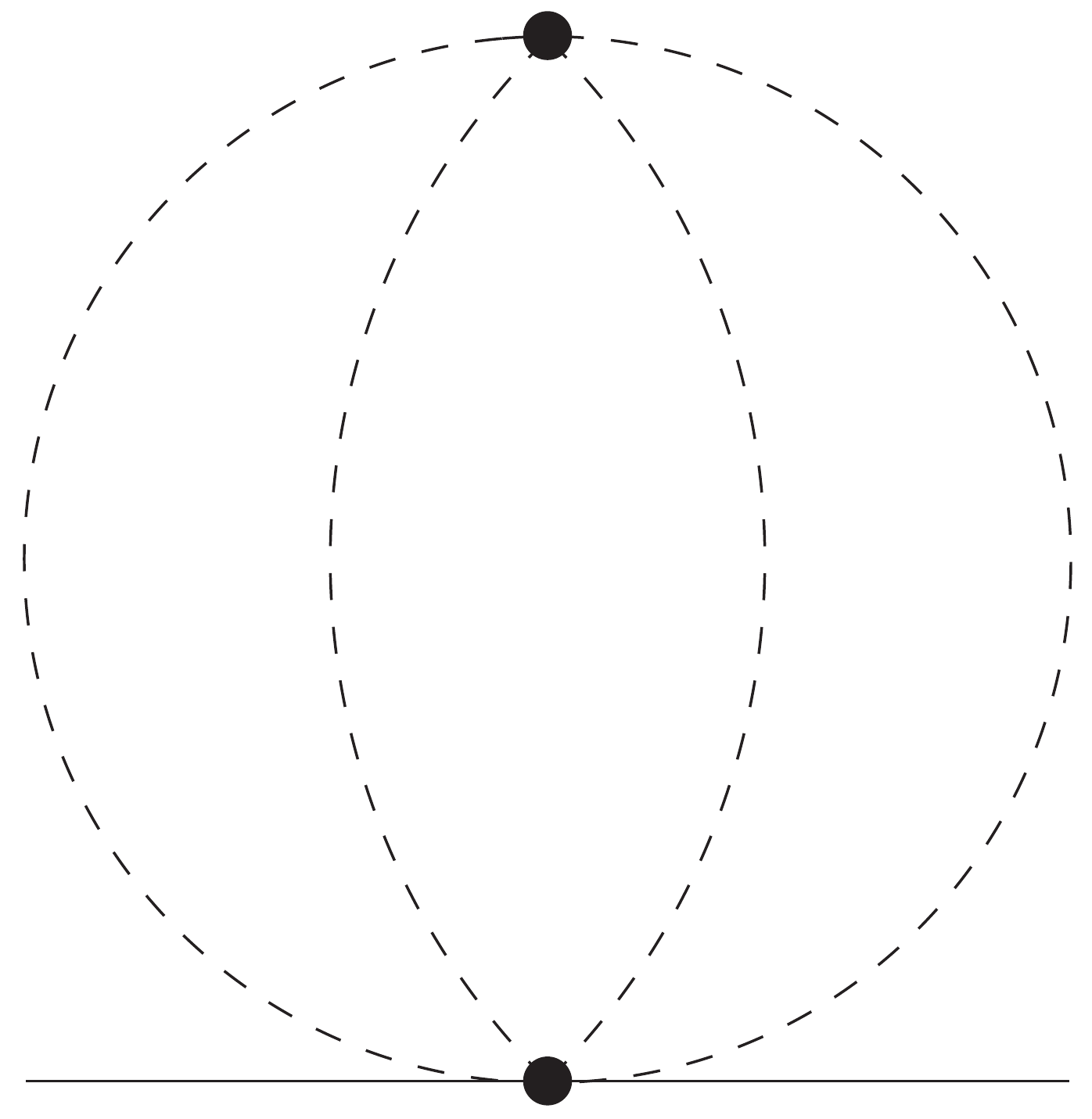}
    \caption*{$M_5$}
  \end{subfigure}
  \begin{subfigure}{0.23\textwidth}
    \centering
    \includegraphics[width=0.55\linewidth]{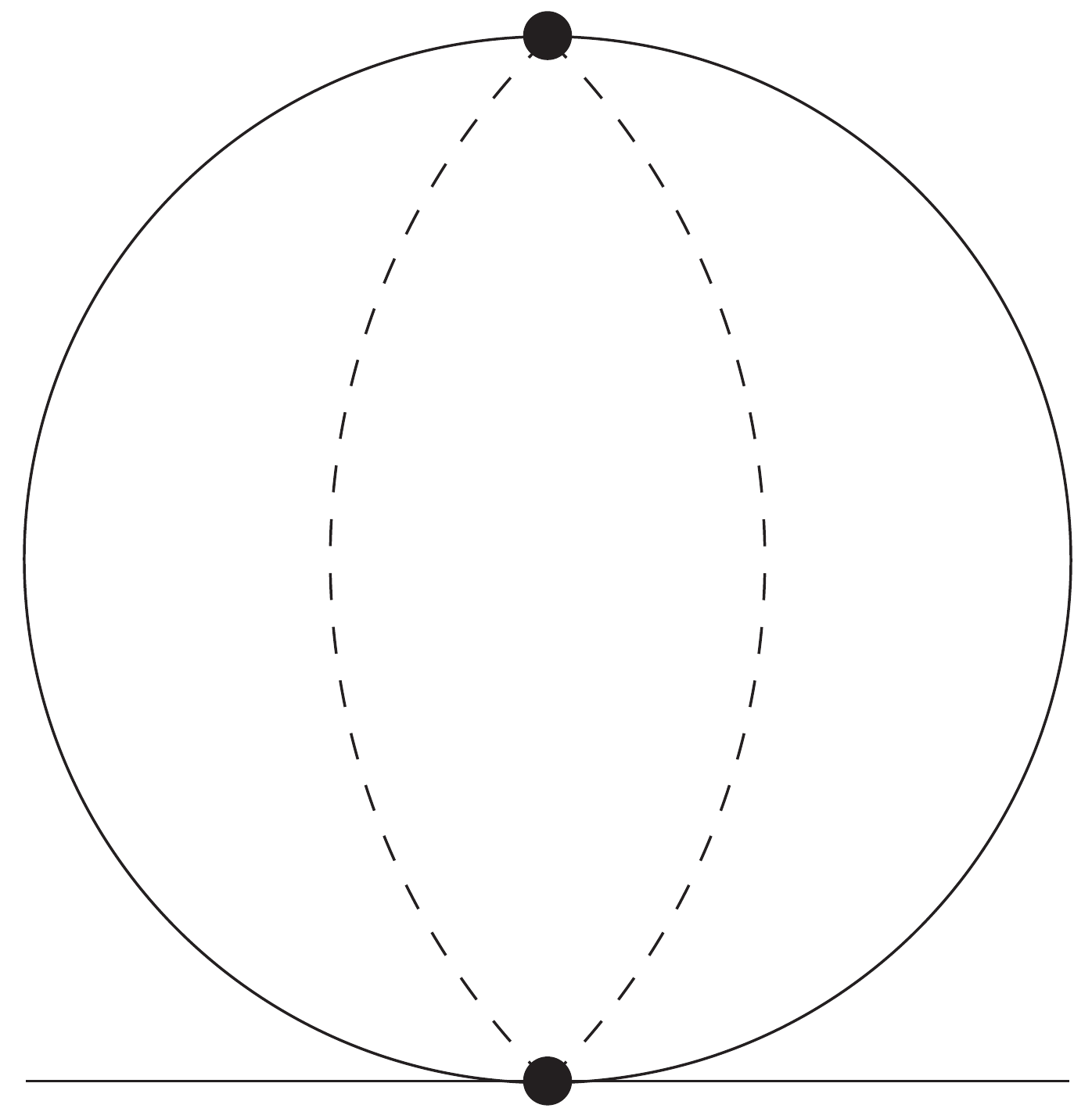}
    \caption*{$M_6$}
  \end{subfigure}
  \begin{subfigure}{0.23\textwidth}
    \centering
    \includegraphics[width=0.55\linewidth]{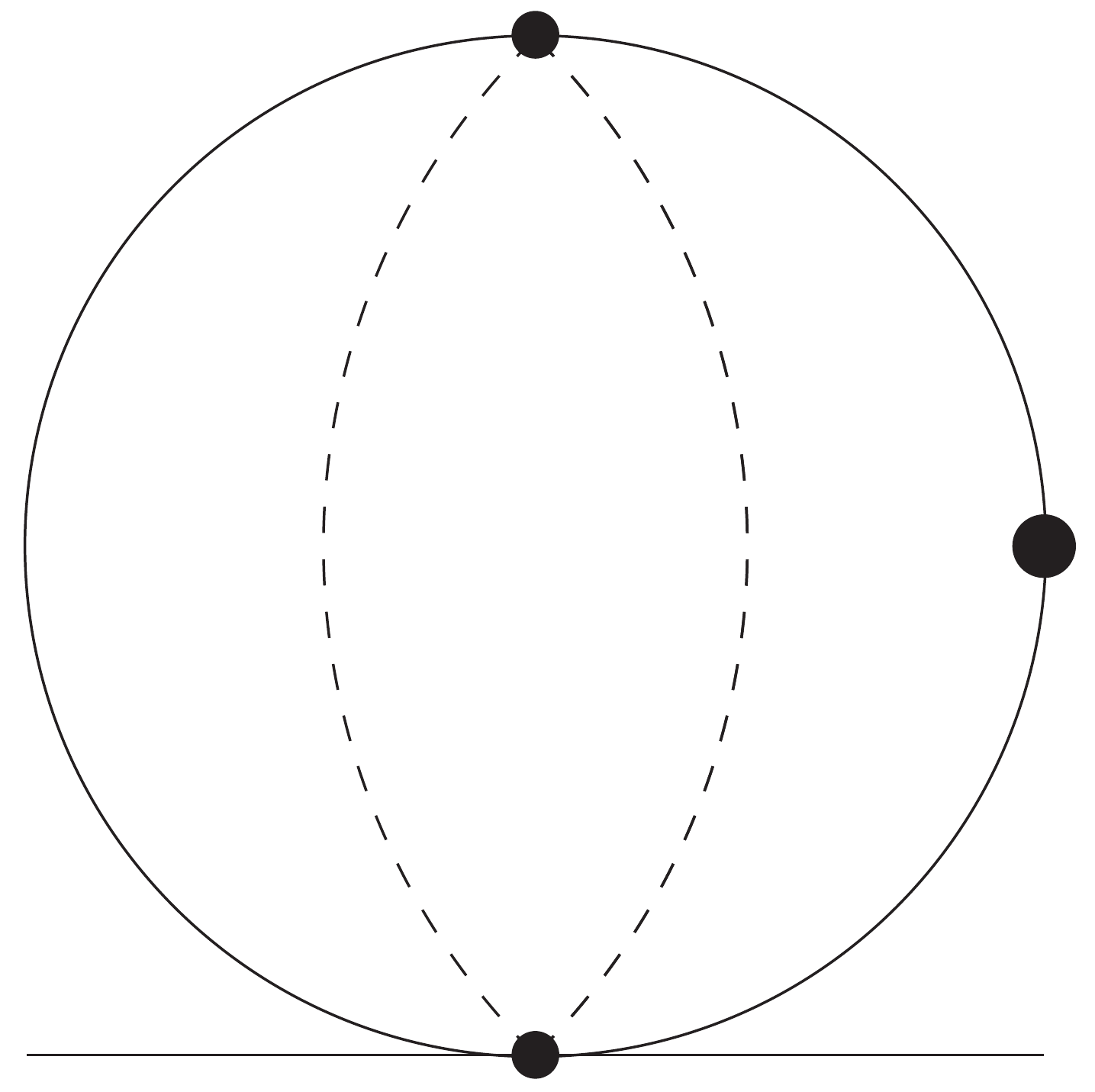}
    \caption*{$M_7$}
  \end{subfigure}
  \begin{subfigure}{0.23\textwidth}
    \centering
    \includegraphics[width=0.83\linewidth]{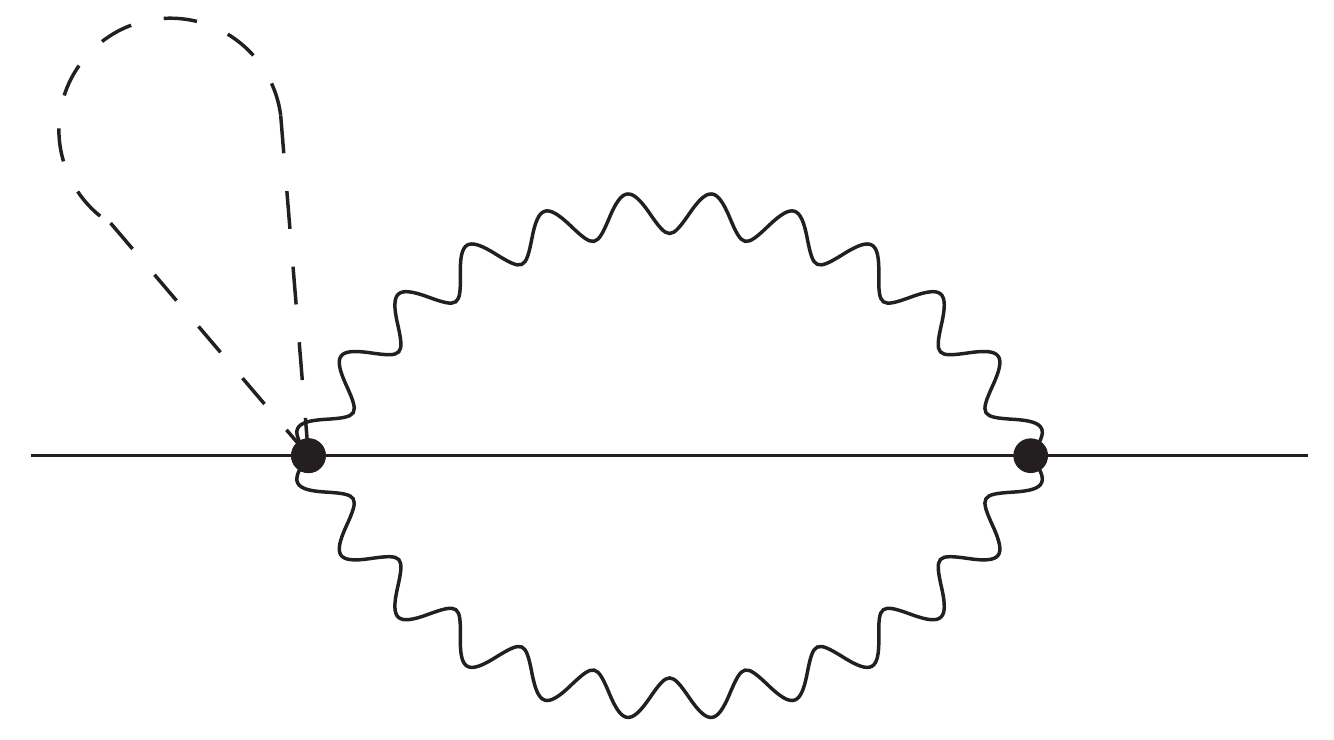}
    \caption*{$M_8$}
  \end{subfigure}
  \begin{subfigure}{0.23\textwidth}
    \centering
    \includegraphics[width=0.83\linewidth]{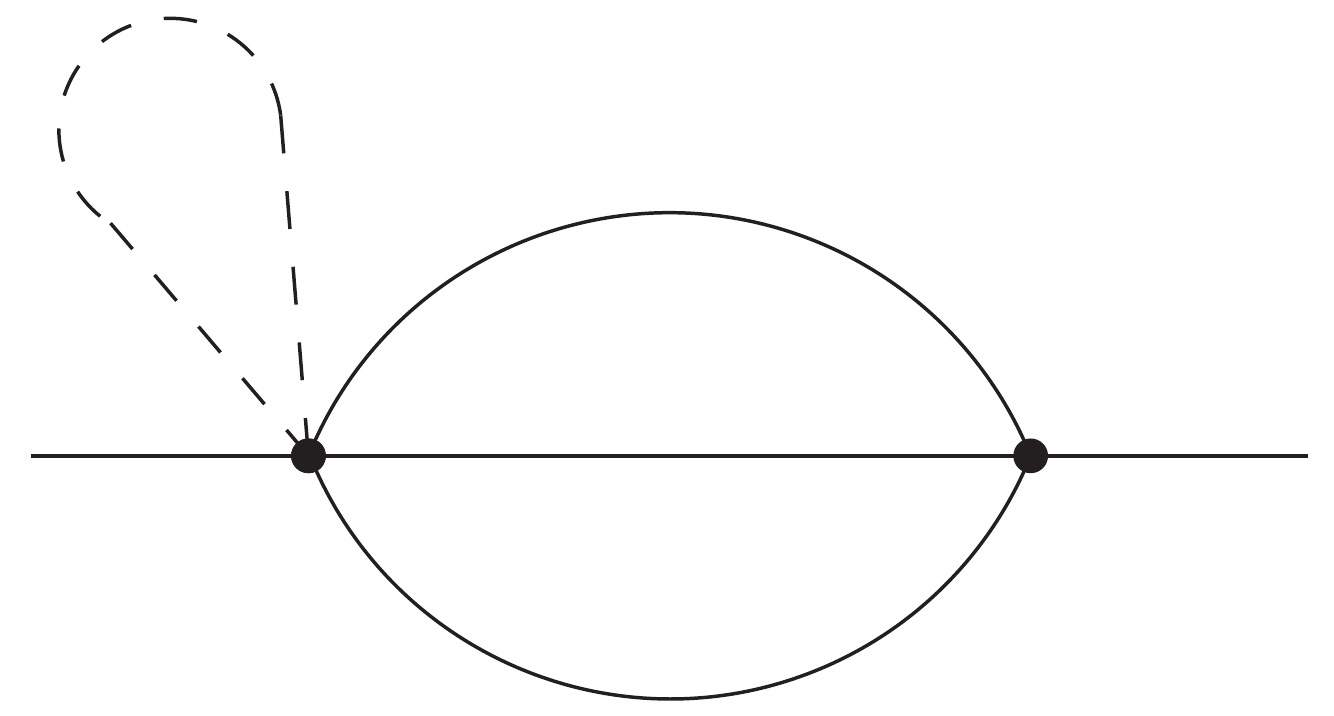}
    \caption*{$M_9$}
  \end{subfigure}
  \begin{subfigure}{0.23\textwidth}
    \centering
    \includegraphics[width=0.83\linewidth]{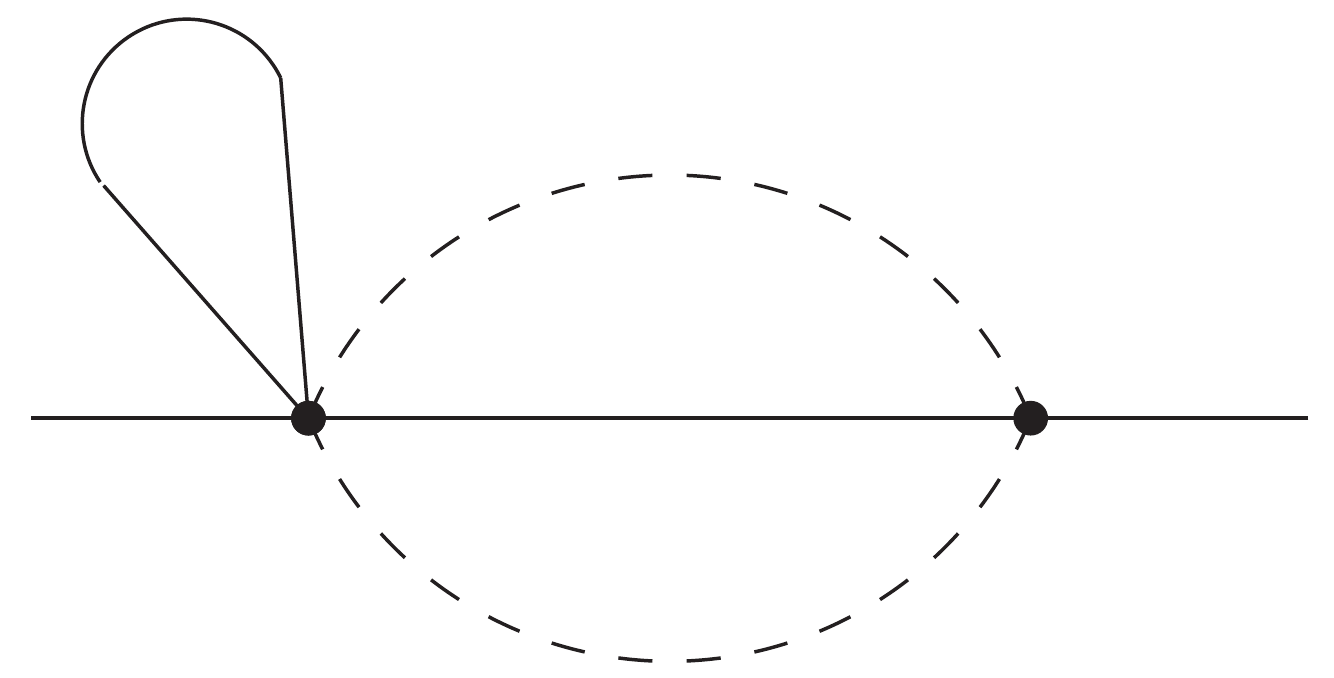}
    \caption*{$M_{10}$}
  \end{subfigure}
  \begin{subfigure}{0.23\textwidth}
    \centering
    \includegraphics[width=0.83\linewidth]{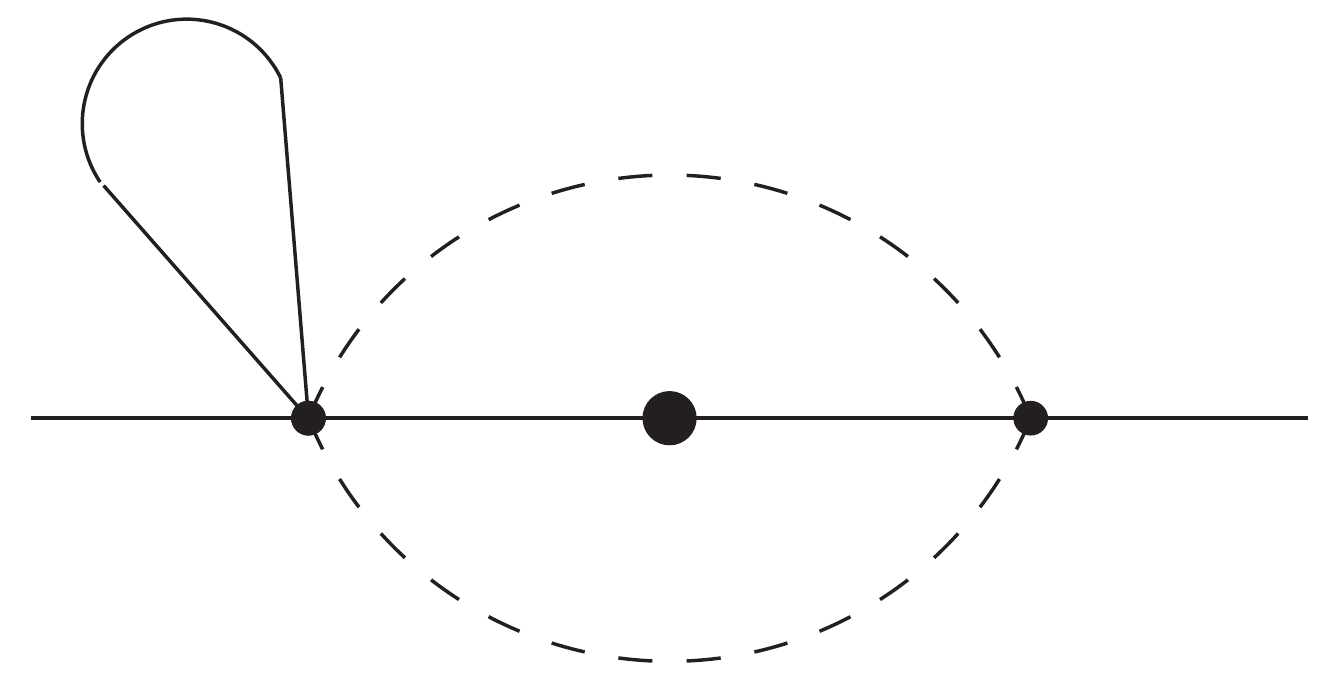}
    \caption*{$M_{11}$}
  \end{subfigure}
    \begin{subfigure}{0.23\textwidth}
    \centering
    \includegraphics[width=0.83\linewidth]{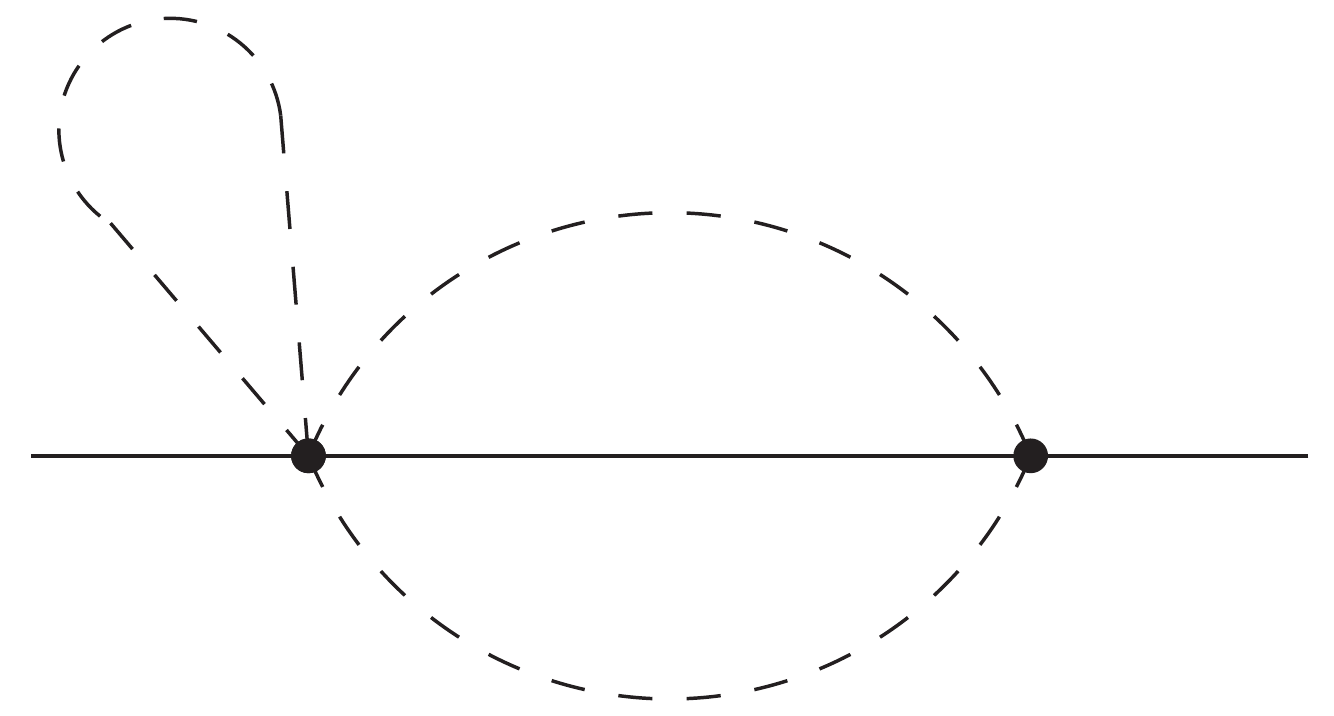}
    \caption*{$M_{12}$}
  \end{subfigure}
  \begin{subfigure}{0.23\textwidth}
    \centering
    \includegraphics[width=0.83\linewidth]{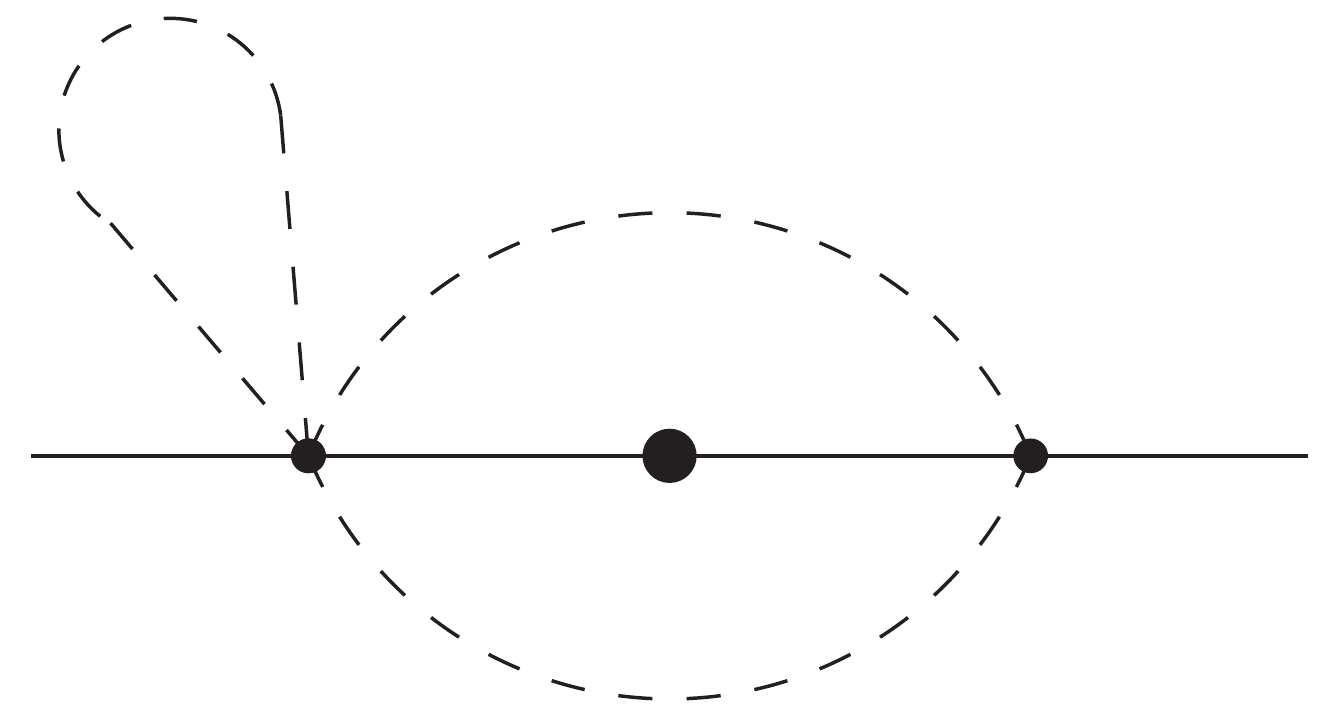}
    \caption*{$M_{13}$}
  \end{subfigure}
  \begin{subfigure}{0.23\textwidth}
    \centering
    \includegraphics[width=0.83\linewidth]{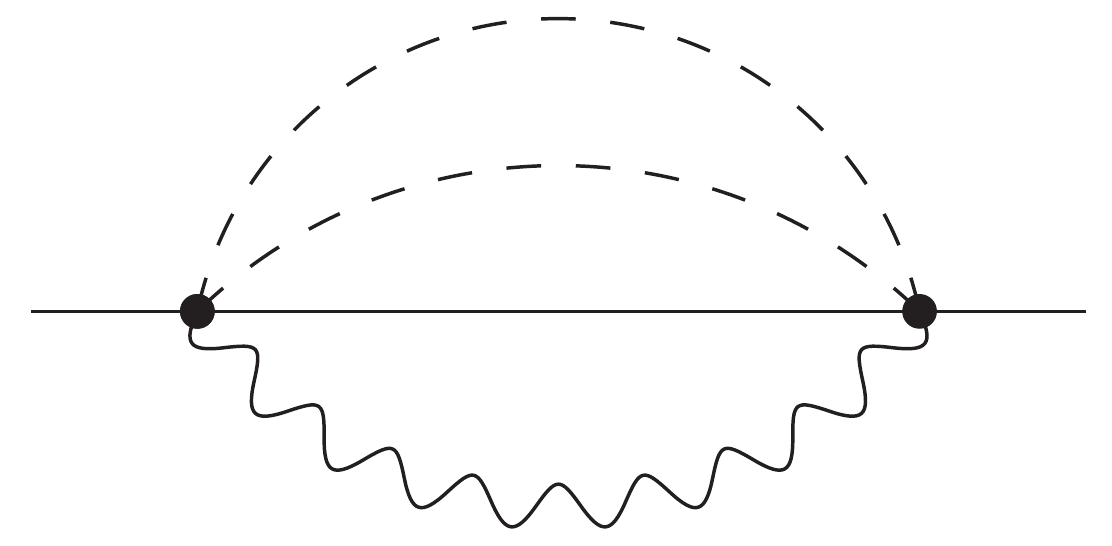}
    \caption*{$M_{14}$}
  \end{subfigure}
  \begin{subfigure}{0.23\textwidth}
    \centering
    \includegraphics[width=0.83\linewidth]{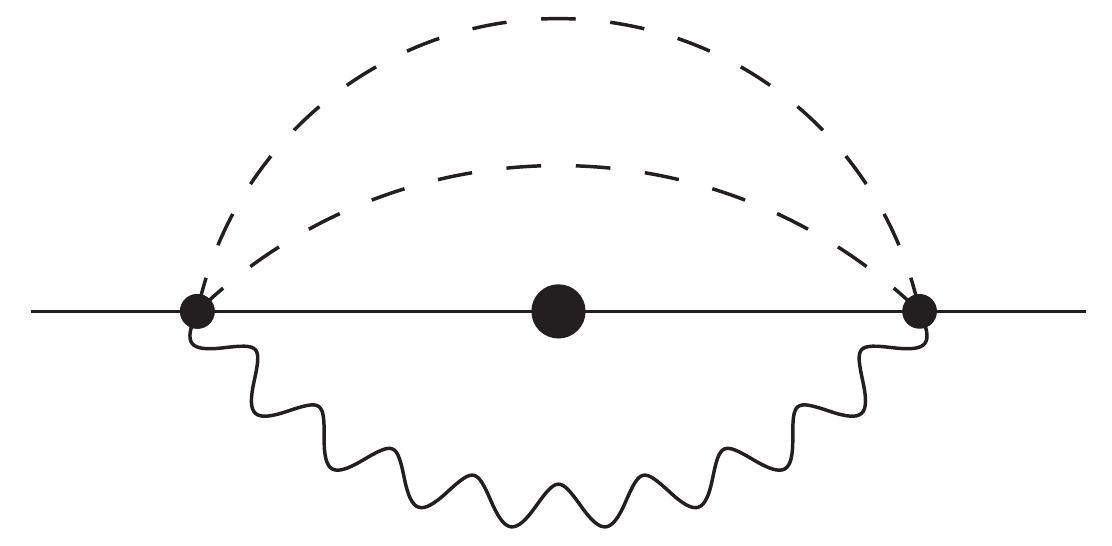}
    \caption*{$M_{15}$}
  \end{subfigure}
    \begin{subfigure}{0.23\textwidth}
    \centering
    \includegraphics[width=0.83\linewidth]{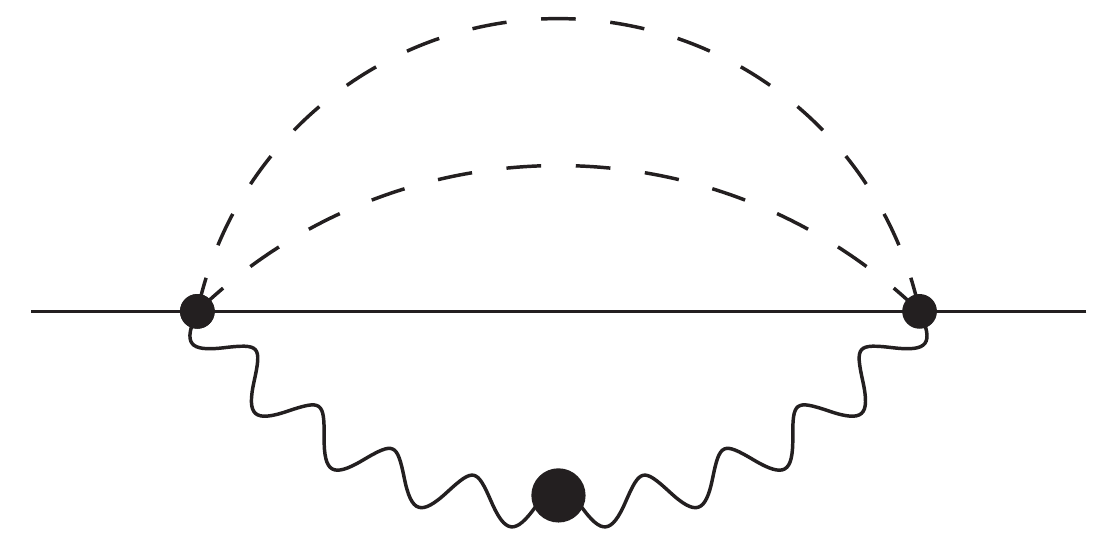}
    \caption*{$M_{16}$}
  \end{subfigure}
  \begin{subfigure}{0.23\textwidth}
    \centering
    \includegraphics[width=0.83\linewidth]{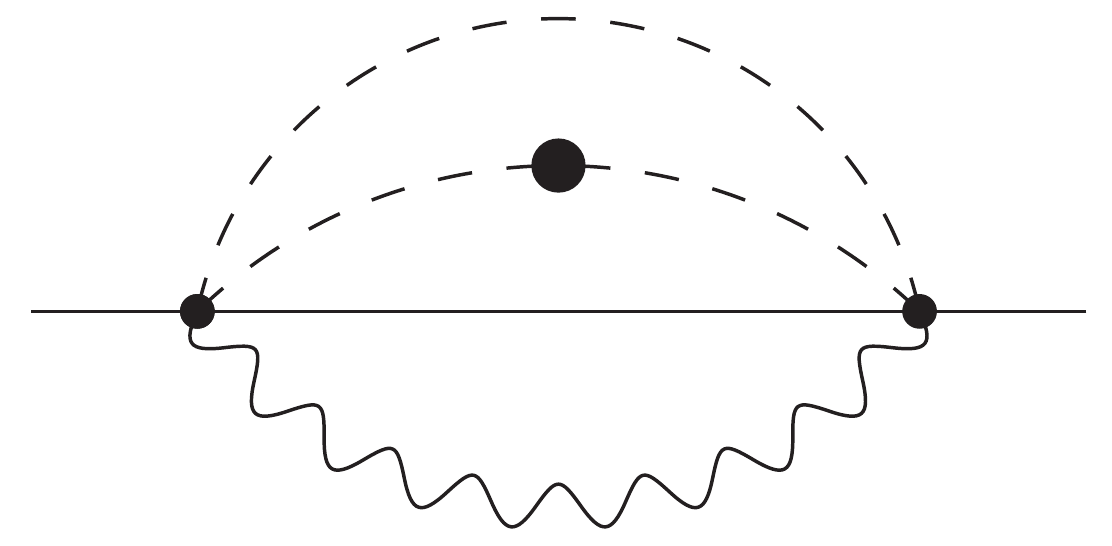}
    \caption*{$M_{17}$}
  \end{subfigure}
  \begin{subfigure}{0.23\textwidth}
    \centering
    \includegraphics[width=0.83\linewidth]{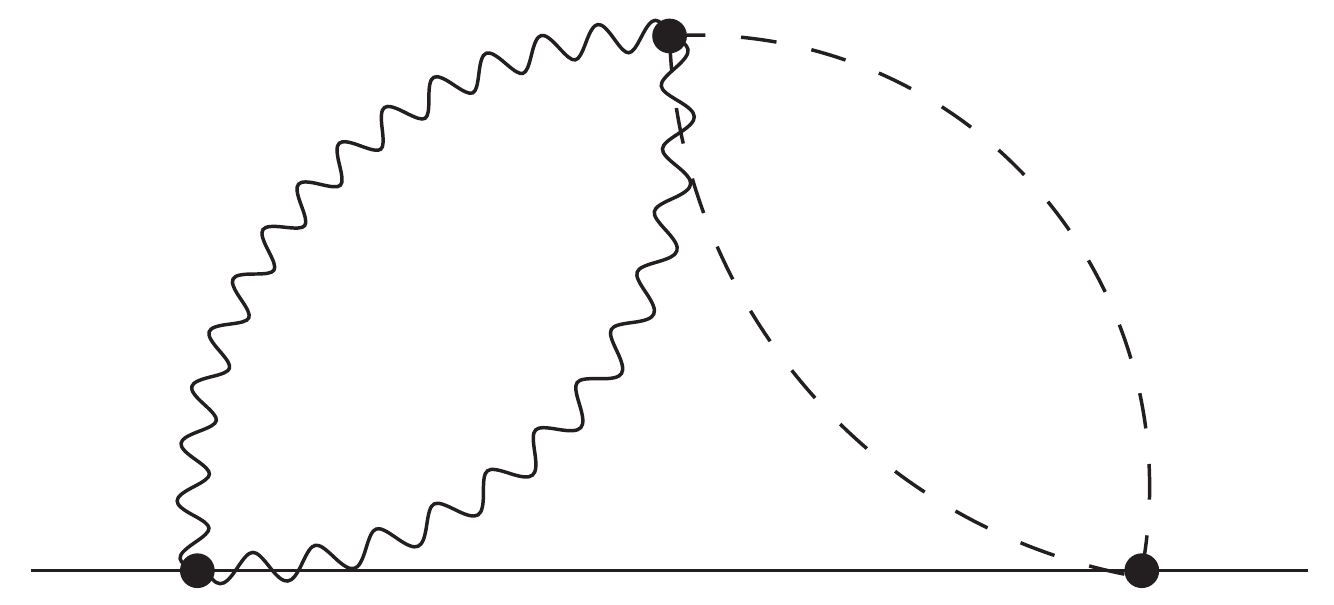}
    \caption*{$M_{18}$}
  \end{subfigure}
  \begin{subfigure}{0.23\textwidth}
    \centering
    \includegraphics[width=0.83\linewidth]{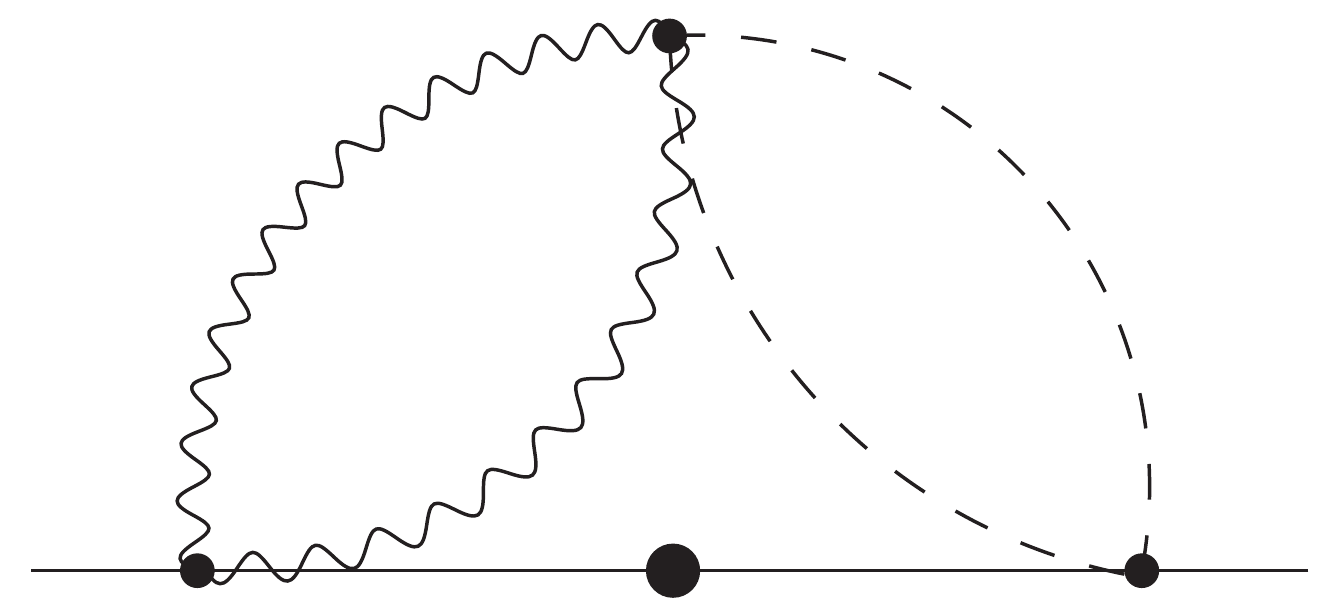}
    \caption*{$M_{19}$}
  \end{subfigure}
    \begin{subfigure}{0.23\textwidth}
    \centering
    \includegraphics[width=0.83\linewidth]{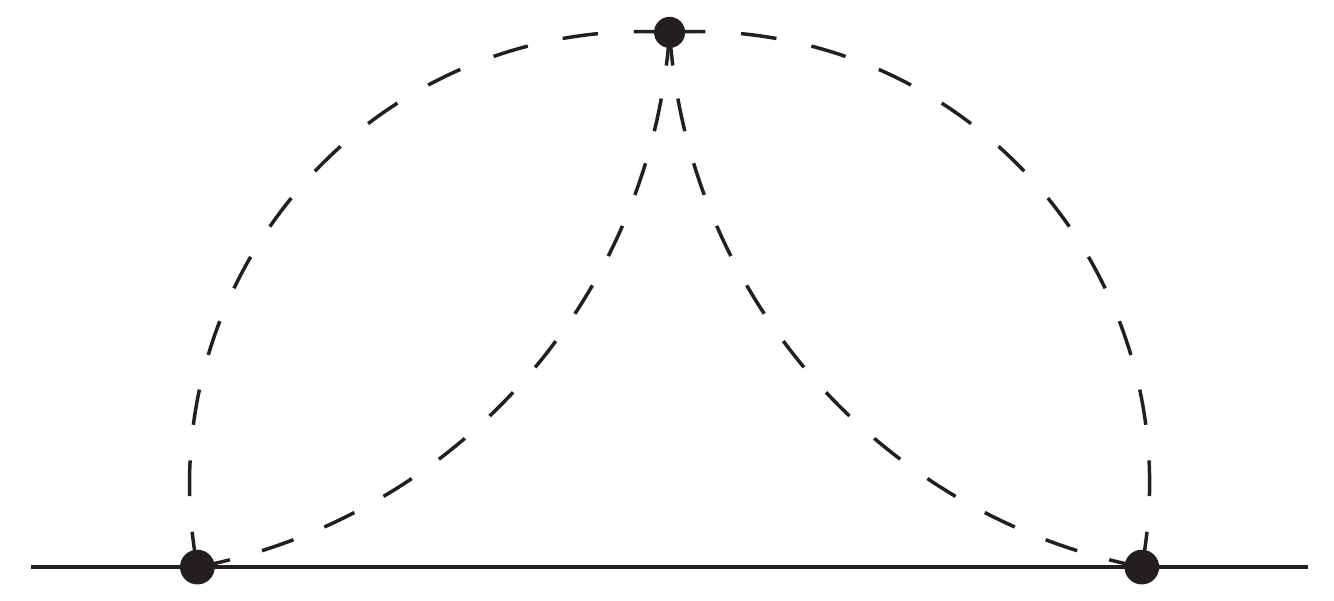}
    \caption*{$M_{20}$}
  \end{subfigure}
  \begin{subfigure}{0.23\textwidth}
    \centering
    \includegraphics[width=0.83\linewidth]{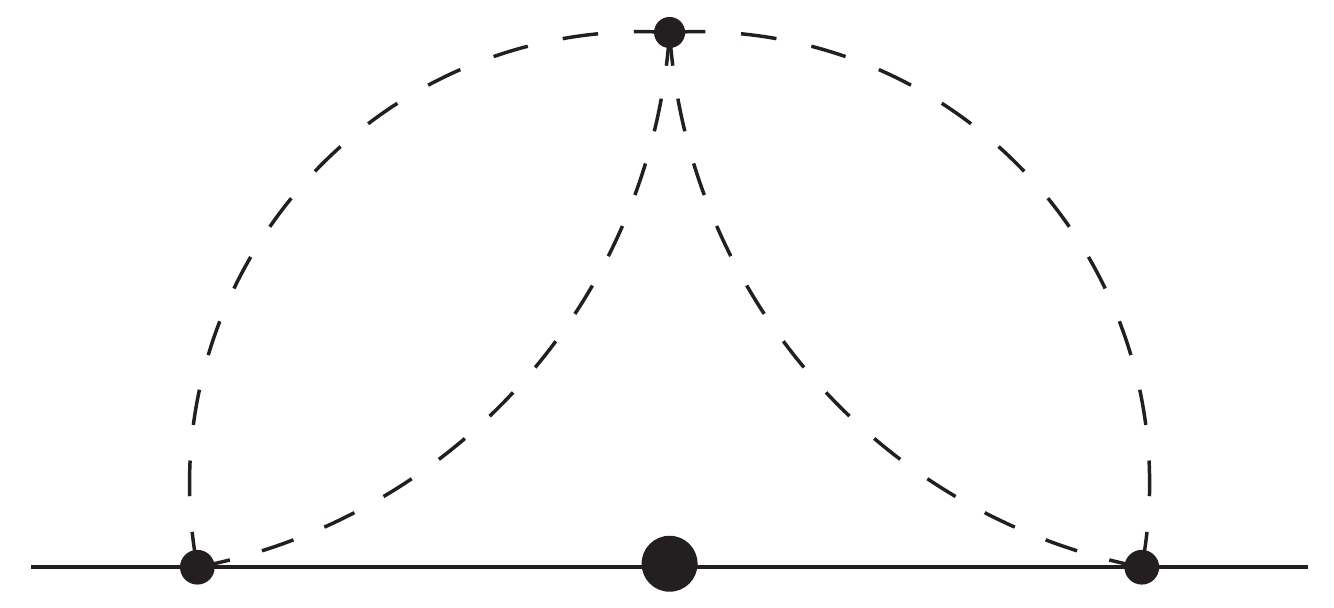}
    \caption*{$M_{21}$}
  \end{subfigure}
  \begin{subfigure}{0.23\textwidth}
    \centering
    \includegraphics[width=0.83\linewidth]{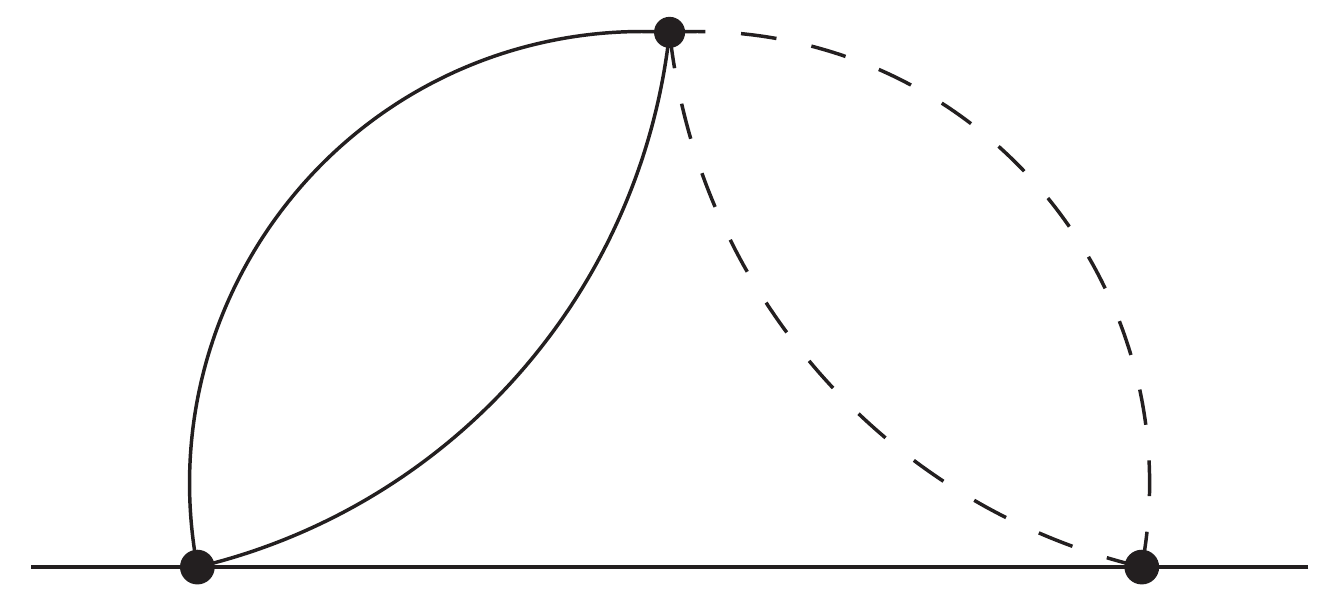}
    \caption*{$M_{22}$}
  \end{subfigure}
  \begin{subfigure}{0.23\textwidth}
    \centering
    \includegraphics[width=0.83\linewidth]{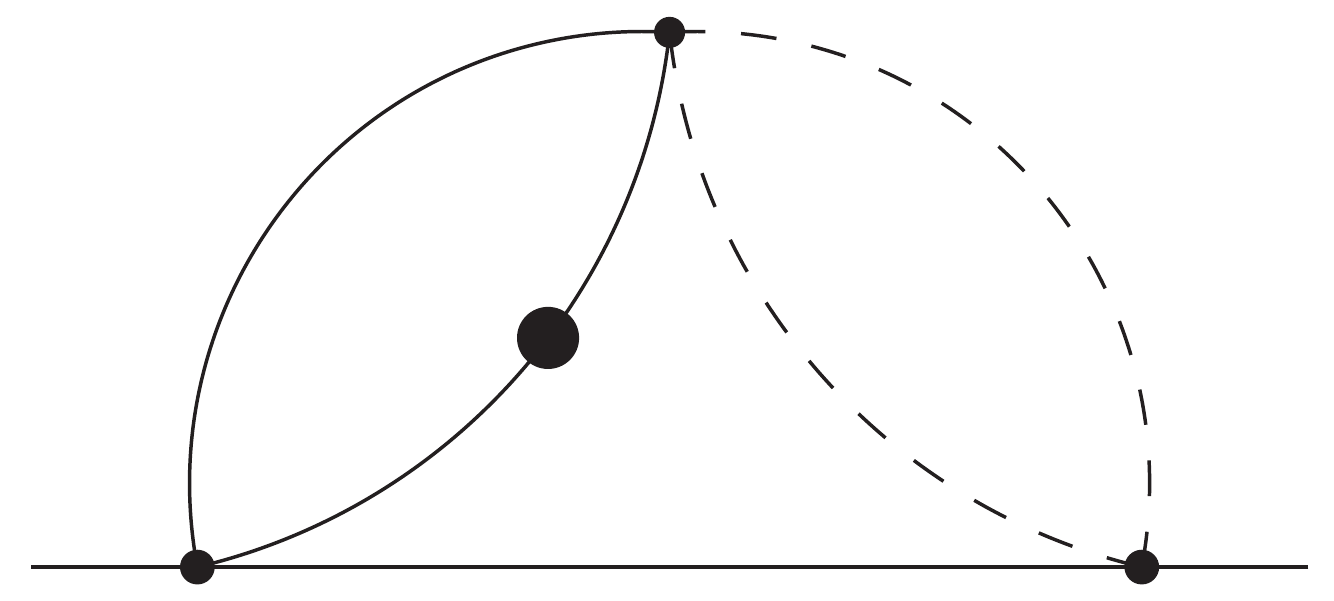}
    \caption*{$M_{23}$}
  \end{subfigure}
  \begin{subfigure}{0.45\textwidth}
    \centering
    \includegraphics[width=0.51\linewidth]{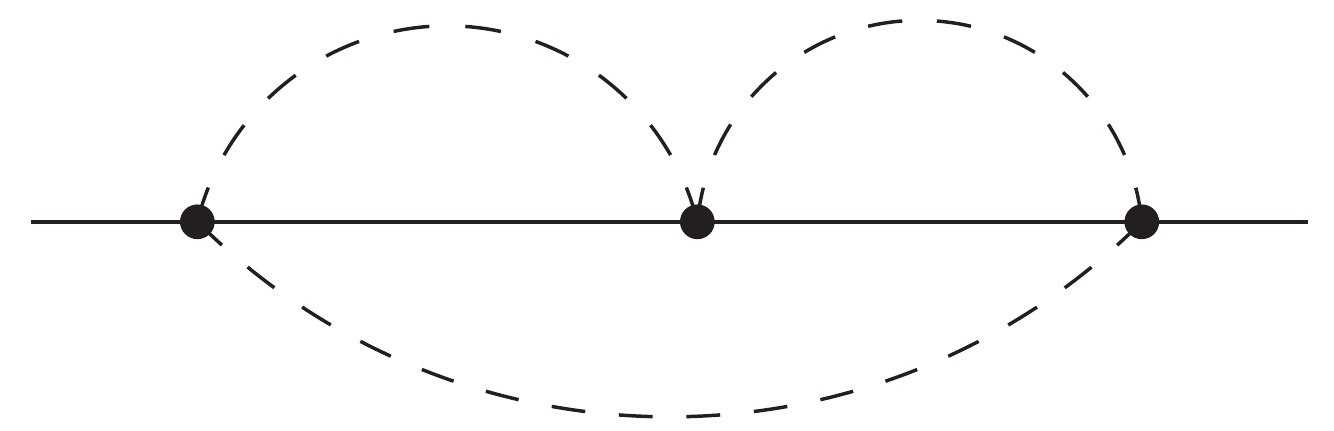}
    \caption*{$M_{24}$}
  \end{subfigure}
  \begin{subfigure}{0.45\textwidth}
    \centering
    \includegraphics[width=0.51\linewidth]{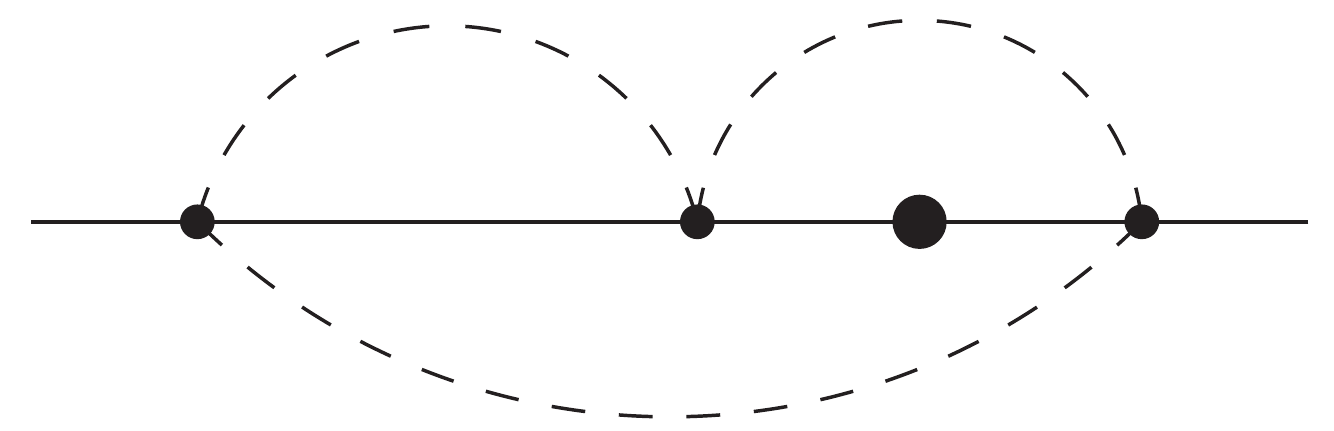}
    \caption*{$M_{25}$}
  \end{subfigure}
  \begin{subfigure}{0.3\textwidth}
    \centering
    \includegraphics[width=0.73\linewidth]{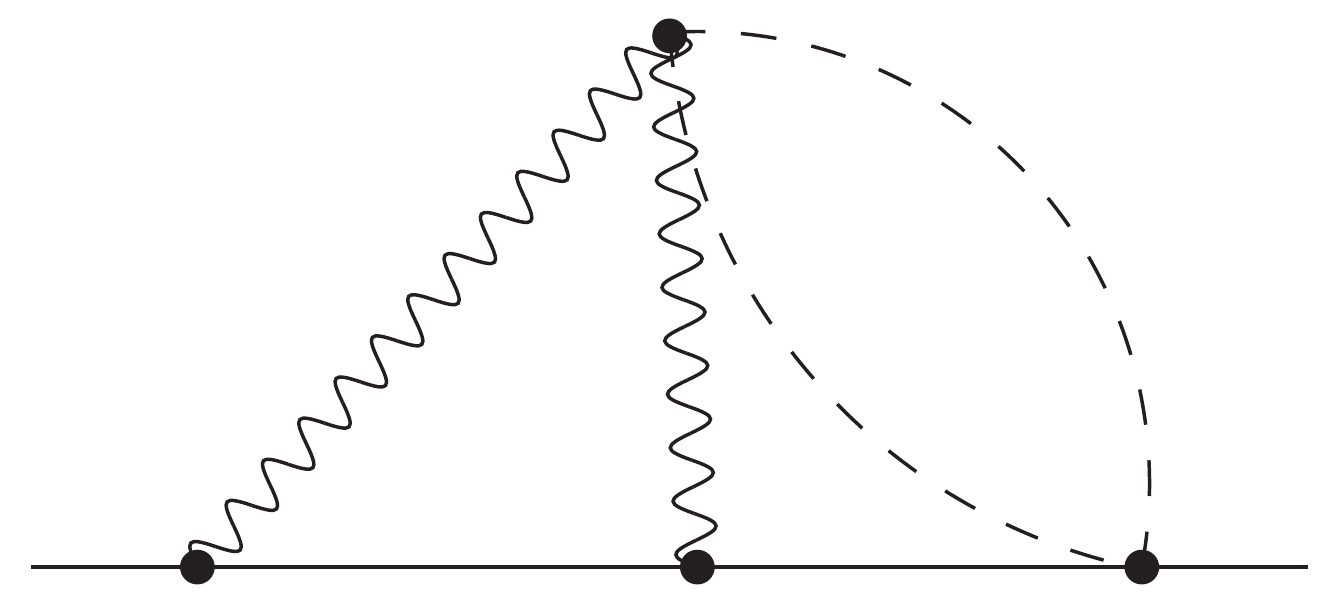}
    \caption*{$M_{26}$}
  \end{subfigure}
  \begin{subfigure}{0.3\textwidth}
    \centering
    \includegraphics[width=0.73\linewidth]{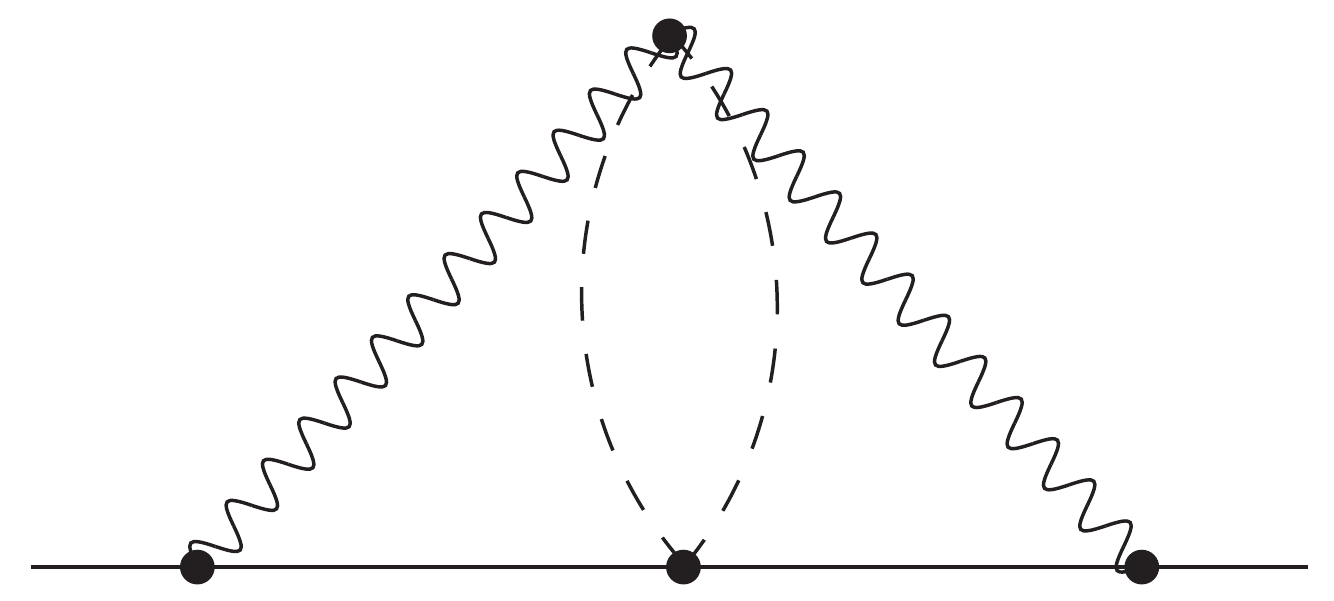}
    \caption*{$M_{27}$}
  \end{subfigure}
  \begin{subfigure}{0.3\textwidth}
    \centering
    \includegraphics[width=0.73\linewidth]{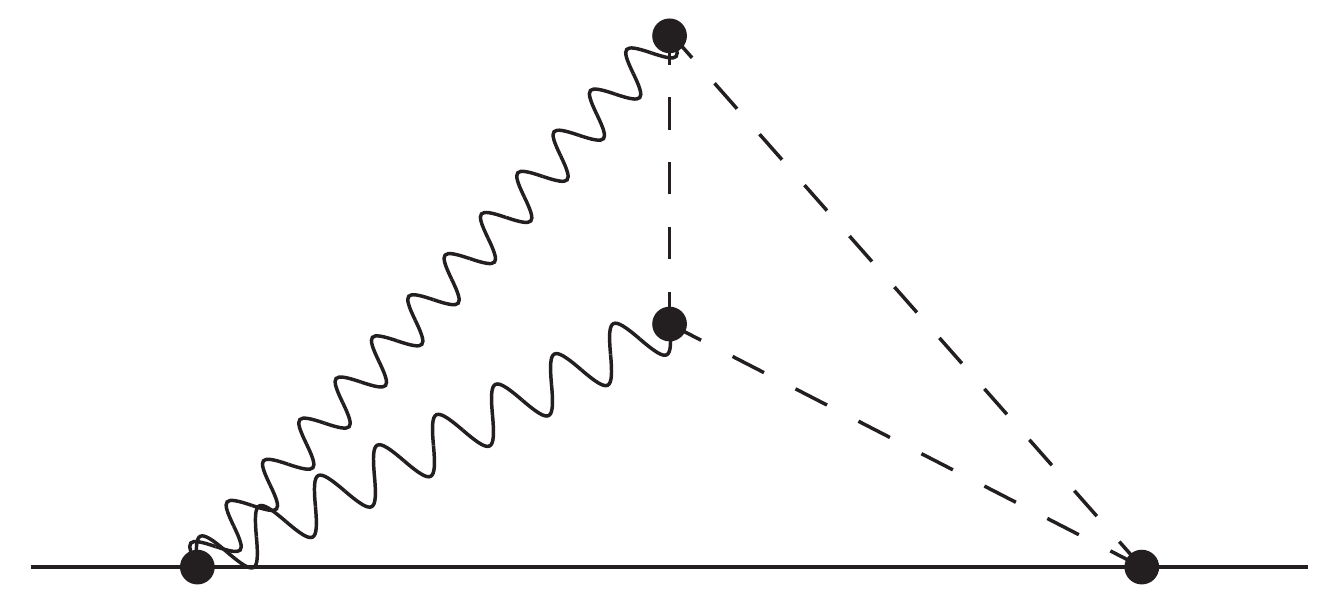}
    \caption*{$M_{28}$}
  \end{subfigure}
  \caption{Three-loop master integrals after the {\tt LiteRed} reduction. 
    Solid and dashed lines represent propagators with mass $m_1$ and $m_2$. Wavy
    lines stand for massless particles. Note that there is a relation between 
    $M_{4}$, $M_{14}$, $M_{15}$ and $M_{17}$ (see text).}
  \label{fig:masters}
\end{figure}

We base our three-loop calculation on intermediate expressions obtained
in Ref.~\cite{Bekavac:2007tk}. In particular, we use the results where
$Z_m^{\rm OS}$ and $Z_2^{\rm OS}$ are expressed in terms of the 28 master
integrals shown in Fig.~\ref{fig:masters}.  They are obtained from three
integral families  introduced
in Refs.~\cite{Bekavac:2007tk,Bekavac:2009gz}.  We have implemented the integral
families in \texttt{LiteRed}~\cite{Lee:2012cn,Lee:2013mka} and redone the
redution to the 28 master integrals shown in Fig.~\ref{fig:masters}.  After
using the additional relation~\cite{Bekavac:2009gz}
\begin{eqnarray}
  M_{17} &=& -\frac{1}{x^2} M_{15}
         - \frac{2d-5}{2x^2} M_{14}
         - \frac{2-d}{4x^2} M_{4} ,
             \label{eq::M17}
\end{eqnarray}
one ends up with 27 master integrals, which have to be computed.

We utilize \texttt{LiteRed}~\cite{Lee:2012cn,Lee:2013mka} to derive a closed
system of differential equations.  Out of the 27 master integrals 23 can be
solved in terms of harmonic polylogarithms (HPLs). Their analytic results are
given in~\cite{Bekavac:2009gz}.  There it was also noted that four
master integrals ($M_{20}$, $M_{21}$, $M_{22}$, $M_{23}$) cannot be expressed
in terms of HPLs at higher orders in $\epsilon = (4-d)/2$.  For these
integrals expansions around $x=0$ were obtained and numerical values for
larger values of $x$ were calculated.  In this work we obtain analytic results
for the missing master integrals and, furthermore, extend all master integrals
by one order in $\epsilon$ such that we obtain the renormalization constants to
$\mathcal{O}(\epsilon^2)$ at two and to $\mathcal{O}(\epsilon)$ at three
loops, respectively. This will be a crucial input for a future four-loop
calculation of the renormalization constants.

We solve the coupled system of differential equations using the algorithmic
approach presented in Ref.~\cite{Ablinger:2018zwz}.  For convenience of the
reader we outline in the following the main steps of this approach.  The
differential equation can be written in the form
\begin{eqnarray}
  \frac{ d \vec{M}(x,\epsilon) }{d x} &=& 
  \mathcal{A}(x,\epsilon) \cdot \vec{M}(x,\epsilon)\,,
\end{eqnarray}
where $\vec{M}(x,\epsilon)$ is the vector of our 27 master integrals.
It can be chosen such that the matrix $\mathcal{A}$ is in
upper-block diagonal form, i.e. the diagonal elements
are square matrices with possible non-vanishing
entries to the left.  

The square matrices on the diagonal represent coupled sets of master
integrals, which only depend on themselves and integrals from lower sectors.
One can therefore solve the system successively starting from simpler systems
and insert the solutions as inhomogeneities into the more involved ones.  In
total we find one $3 \times 3$ and seven $2 \times 2$ systems.  The $x$
dependence of seven of the remaining ten master integrals factorizes. 

We decouple the systems of differential equations with the package
\texttt{OreSys}~\cite{ORESYS}, which is based on
\texttt{Sigma}~\cite{Schneider:2007}, to obtain a single 
differential equation of higher order for one of the master integrals in the
system. Furthemore, \texttt{OreSys} provides rules
to construct the other master integrals from the solution of the differential
equation.  The higher order
differential equations are then expanded in $\epsilon$ and iteratively solved
order by order.  To solve the differential equations we make use of the solver
implemented in \texttt{HarmonicSums}~\cite{HarmonicSums}, which is
particularly well suited to find solutions in terms of iterated integrals.  In
a first step we consider the homogeneous part of the differential equation and
try to write it in factorized form.  If it fully factorizes into first order
factors the solution in terms of iterated integrals can be obtained in a
straightforward way.  If second order factors remain Kovacic's algorithm~\cite{Kovacic}
is used to find all solutions of the differential equation, which can be written
in terms of iterated integrals.  In our case the homogenous solutions
in terms of iterated integrals exist and thus also the particular solutions can
be expressed in terms of iterated integrals.  The
construction and simplification of the homogeneous and particular solution is
automated in \texttt{HarmonicSums}.  To fully solve the differential equations
we still need to fix the integration constants multiplying the homogeneous
solutions.  Boundary values for all integrals at $x=0$ and $x=1$ can be
extracted from the on-shell integrals given in Ref.~\cite{Lee:2010ik}.  To fix
all integration constants we need both limits since for some master integrals
the homogeneous solutions vanish at $x=0$ or $x=1$.

Four master integrals cannot be expressed in terms of usual HPLs.
We want to illustrate this for the system of differential
equations of the integrals $M_{22}$ and $M_{23}$.
After decoupling the homogenous differential equations
for the master integral $M_{22}$ we obtain
\begin{eqnarray}
  M_{22}^{\prime \prime} &=& 
    \frac{1-4x^2}{x(1-x^2)} M_{22}^{\prime}
  + \frac{4}{1-x^2} M_{22}\,,
                             \label{eq::M22}
\end{eqnarray}
where $d=4$ has been used in the coefficients since the $\epsilon$-dependent
terms enter the inhomogenious part. Equation~(\ref{eq::M22}) has the two solutions
\begin{eqnarray}
  M_{22}^{(1)} &=& x^2(4-x^2)\,,
  \nonumber\\
  M_{22}^{(2)} &=& (2-3x^2+x^4) \sqrt{1-x^2} + x^2(4-x^2) \left[ 1 +
                   I\left(\left\{\frac{\sqrt{1-\tau^2}}{\tau}\right\}
                   ,x\right)\right]\,, 
                   \label{eq::M22_sol}
\end{eqnarray}
where $I$ denotes a generalized iterated integral over the
specified integration kernels, i.e.
\begin{eqnarray}
  I\left(\left\{ g(\tau) , \vec{h}(\tau) \right\} , x \right) 
  &=&
      \int\limits_{0}^{x} {\rm d}t\, g(t)\, I\left(\left\{  \vec{h}(\tau) \right\} , t \right)\,.
      \label{eq::it_int}
\end{eqnarray}
Note that a regularization is needed for letters which lead to divergent
expressions for $t\to 0$. This is in complete analogy to 
HPLs~\cite{Remiddi:1999ew}.  Equation~(\ref{eq::M22_sol})  illustrates that one has to
introduce the new letter $\sqrt{1-\tau^2} / \tau$ in order to solve the
differential equation.  Analogously the system of master integrals $M_{20}$
and $M_{21}$ introduces the letter $\sqrt{1-\tau^2}$.  After fixing the
boundary conditions it turns out that for all master integrals the generalized
letters are only needed from $\mathcal{O}(\epsilon)$ onwards.
Note that the ${\cal O}(\epsilon)$ terms enter the finite
contribution of $Z_m^{\rm OS}$ and $Z_2^{\rm OS}$.

Since the additional letters only introduce one square root 
it is possible to rationalize the letters with a suitable
variable transformation.
One possibility is the so-called trigonometric substitution
\begin{eqnarray}
  x &=& \frac{2y}{1+y^2} \,,
  \label{eq:subs}
\end{eqnarray}
which introduces the letters
\begin{eqnarray}
  \frac{1}{1+\tau^2}\,,\quad\frac{\tau}{1+\tau^2}\,.
\end{eqnarray}
Iterated integrals over these kinds of letters have been studied in
Ref.~\cite{Ablinger:2011te}; the corresponding iterated integrals are
called cyclotomic HPLs.

Alternatively one can factor the polynomial over the
complex numbers and introduce Goncharov polylogarithms~\cite{Goncharov:1998kja}
with letters taken from the set of the 4th root of unity.
For example one has
\begin{eqnarray}
  I\left(\left\{\sqrt{1-\tau^2}\right\} ,x\right) &=&
  \frac{y(1-y^2)}{(1+y^2)^2} + I\left(\left\{\frac{1}{1+\tau^2}\right\},y\right)
  \nonumber\\
  &=& \frac{y(1-y^2)}{(1+y^2)^2} + \frac{i}{2} \left[ G(i,y) - G(-i,y) \right]
  ~,
\end{eqnarray}
but also 
\begin{eqnarray}
  H_1(x) &=& 2H_1(y)+2 I\left(\left\{\frac{\tau}{1+\tau^2}\right\},y\right)\,,
\end{eqnarray}
which shows that the variable transformation in Eq.~(\ref{eq:subs}) converts
HPLs with argument $x$ into cyclotomic HPLs with argument $y$.  Note, however,
that the transformation in Eq.~(\ref{eq:subs}) significantly increases the
complexity of the rational functions in the differential equations. Thus, we
have chosen to solve them in the variable $x$.  However, Eq.~(\ref{eq:subs})
is needed to fix the boundary conditions at $x=1$, since this requires the
evaluation of the iterated integrals at this point. The corresponding results
up to weight~5 are conveniently obtained by transforming the iterated
integrals to cyclotomic HPLs for which the values at $x=1$ are known up to
weight~6~\cite{Ablinger:2011te}.  This leads to relations like
\begin{eqnarray}
  I\left(\left\{\frac{1}{\tau},\sqrt{1-\tau^2},\frac{1}{1-\tau}\right\}
  ,1\right) 
  &=&
  \frac{5}{4}
  + C 
  \biggl(
    \frac{1}{2}
    +l_2
  \biggr)
  -\frac{3 l_2}{2} 
  +\frac{l_2^2 \pi }{4}
  -\frac{\pi ^2}{8}
  -\frac{\pi ^3}{96}
  + \frac{1}{2} cs_1 ~,
  \nonumber\\
\end{eqnarray}
with $l_2=\log(2)$, Catalan's constant 
\begin{eqnarray}
  C &=& \sum_{i=1}^{\infty} \frac{(-1)^i}{(2i+1)^2} \,\,\approx\,\, 0.915966\,,
  \label{eq::Catalan}
\end{eqnarray}
and a further cyclotomic constant
\begin{eqnarray}
  cs_1 &=& \sum\limits_{n=1}^{\infty} \frac{(-1)^n}{n^2} 
           \sum\limits_{m=1}^{n} \frac{(-1)^m}{1+2m} 
        \,\,\approx\,\, 0.330798\,.
\end{eqnarray}
Note that for the evaluation of individual functions at argument $x=1$
several cyclotomic constants appear. However, in our final result,
all but $C$ cancel.
Analytic results for all master integrals are provided in an
ancillary file to this paper~\cite{progdata}.

For fast and precise numerical evaluations we provide expansions around $x=0$,
$x=1$ and $x\to \infty$.  The expansions around $x=0$ can easily be
obtained utilizing \texttt{HarmonicSums}.  For the expansion around $x=1$ we
first map the argument of the iterated integrals to $1-x$.  This can be
achieved iteratively with the formula
\begin{eqnarray}
  I\left( \left\{ w_1(\tau),\ldots,w_n(\tau) \right\} ,x\right) 
  &=& I\left( \left\{ w_1(\tau),\ldots,w_n(\tau) \right\} ,1\right)
      \nonumber \\ 
  && - \int\limits_{0}^{1-x} {\rm d}t w_{1}(1-t) I\left( \left\{
     w_2(\tau),\ldots,w_n(\tau) \right\} ,1-t\right)\,, 
\end{eqnarray}
which can be easily proven from the integral representation.  In our case this
step does not introduce new letters, but introduces the iterated integrals at
argument $x=1$.  The same constants were already needed to fix boundary
conditions for the differential equations.  Afterwards we can expand easily
around $1-x$.

The expansion for $x\to \infty$ is more involved since the letters involving
square roots develop a brach cut for $x>1$. Thus, in a first step we have to
construct the analytic continuation for the iterated integrals, i.e., the
relations for the corresponding functions with argument $x<1$.  We
use differential equations to do this. Let us for illustration consider an
iterated integral of weight one. Then we have
\begin{eqnarray}
  \frac{d}{dx} I \left( \left\{ \frac{\sqrt{1-\tau^2}}{\tau} \right\},x \right) 
  &=& \frac{\sqrt{1-x^2}}{x}\,.
      \label{eq::dI}
\end{eqnarray}
Now we change the variable to $z=1/x$ and find
\begin{eqnarray}
  \frac{d}{dz} f(z) &=& - i \frac{\sqrt{1-z^2}}{z^2}\,,
\end{eqnarray}
where $f(z)$ is the analytic continuation of the iterated integral in
Eq.~(\ref{eq::dI}).  We assume $0<z<1$ in accordance with $x>1$.  Note that in
our case the change of variables again does not introduce new letters.  The
differential equation can be easily solved by integrating the right hand side
over $z$ and fixing the integration constant for $x=z=1$.  This again only
requires the knowledge of the iterated integrals at argument $x=1$. For our
example, we obtain
\begin{eqnarray}
  I \left( \left\{ \frac{\sqrt{1-\tau^2}}{\tau} \right\},x \right) 
  &=& l_2 - 1 + i \left[ \frac{(1-z^2)^{3/2}}{z} + 2 I \left( \left\{\sqrt{1-\tau^2}\right\},z \right) - \frac{\pi}{2} \right]~.
      \label{eq::I_ana_cont}
\end{eqnarray}
For higher weights one can proceed iteratively, since the derivative of 
an iterated integral of weight $w$ with respect to its argument  only 
depends on iterated integrals of weight $w-1$.

Note that the analytic continuation of the individual iterated integrals
introduces imaginary parts (cf. Eq.~(\ref{eq::I_ana_cont})). However, after
inserting the analytic continuations for all iterated integrals into the
expressions for $Z_m^{\rm OS}$ and $Z_2^{\rm OS}$ all imaginary parts
cancel analytically and the expansion around $1/x=z=0$ can be obtained in a
straightforward way.


\section{\label{sec::res}Results and conclusions}

In this section we briefly discuss our results for $Z_m^{\rm OS}$, $z_m$ and
$Z_2^{\rm OS}$. After inserting the exact master integrals into the
corresponding amplitudes we renormalize the quark masses $m_1$ and $m_2$ in
the on-shell scheme, the strong coupling constant in the $\overline{\rm MS}$
scheme and expand in $\epsilon$ such that we obtain results up to $\epsilon^2$
at two-loop and $\epsilon^1$ at three-loop order. Whereas the two-loop
results are still quite compact (see, e.g., Eqs.~(15) and~(28) of
Ref.~\cite{Bekavac:2007tk}), the three-loop expressions are too big to be
printed. Instead we provide the analytic expressions in the ancillary files to
this paper~\cite{progdata}.  We also provide transformation rules which map
the iterated integrals introduced in the previous section to Goncharov
polylogarithms which can be evaluated numerically with the help of {\tt
  GiNaC}~\cite{Bauer:2000cp}.  Note that our final three-loop result contains
iterated integrals up to weight five and six in the $\epsilon^0$ and
$\epsilon^1$ term.

More compact expressions are obtained after expanding for $x\to0$, $x\to1$ or
$x\to \infty$. For illustration we show for the $n_m$ dependent terms of $z_m$,
which we define via
\begin{eqnarray}
  z_m^M &=& z_m - z_m(n_m=0)\,,
\end{eqnarray}
the first three expansion
terms at two and three loops.  To keep the expressions compact we specify the
colour factors to QCD, i.e. $C_A=3$, $C_F=4/3$ and $T_F=1/2$.  Furthermore we
set $\mu=m_1$, $n_h=1$ and restrict ourselves to the $\epsilon^0$ term. For
$x\to0$ we obtain
\begin{eqnarray}
  z_m^M &=&
  \left(\frac{\alpha_s}{\pi}\right)^2
  \biggl[
    \frac{71}{144}+\frac{\pi ^2}{18}-\frac{\pi ^2 x}{6}+x^2
  \biggr]
  \nonumber \\ &&
  + \left(\frac{\alpha_s}{\pi}\right)^3
  \biggl[
    \frac{40715}{3888}
    +\frac{941 \pi ^2}{648}
    -\frac{61 \pi ^4}{1944}
    +\frac{695}{216} \zeta_3
    -\frac{8 a_4}{27}
    +\frac{11 \pi ^2 l_2}{81}
    -\frac{2}{81} \pi ^2 l_2^2
    \nonumber \\ &&
    -\frac{l_2^4}{81}
    + n_l 
    \biggl(
      -\frac{2353}{11664}
      -\frac{13 \pi ^2}{162}
    -\frac{7}{27} \zeta_3
      -\frac{2 x^2}{9}
      +x 
      \Big(
        \frac{7 \pi ^2}{27}
        -\frac{2}{9} \pi ^2 l_2
        -\frac{1}{9} \pi ^2 l_x
      \Big)     
    \biggr)
    \nonumber \\ &&
    +x \biggl(
      -\frac{29513 \pi ^2}{2430}
      -\frac{13 \pi ^3}{162}
      +\frac{1199 \pi ^2 l_2}{81}
      +\frac{31 \pi ^2 l_x}{18}
    \biggr)
    +x^2 \biggl(
      \frac{62}{9}
      +\frac{13 \pi ^2}{12}
      \nonumber \\ &&
      +\Big(
        4
        +\frac{3 \pi ^2}{2}
        -\frac{\pi ^4}{12}
      \Big) l_x
      +\frac{11}{2} \zeta_3
      -\frac{3}{4} \pi ^2 \zeta_3
      -\frac{5}{2} \zeta_5
    \biggr)
  \biggr]
  + \mathcal{O}(\epsilon,x^3,\alpha_s^4)
                      \,,
\end{eqnarray}
with $l_x=\log(x)$, $a_4=\mbox{Li}_4(1/2)$ and $\zeta_n$ is Riemann's zeta function.
For $x\to1$ we have
\begin{eqnarray}
  z_m^M &=&
  \left(\frac{\alpha_s}{\pi}\right)^2
  \biggl[
    \frac{143}{144}
    -\frac{\pi ^2}{9}
    +\biggl(
      -\frac{4}{3}
      +\frac{2 \pi ^2}{9}
    \biggr) y
    +\biggl(
        1
        -\frac{\pi ^2}{12}
    \biggr) y^2
  \biggr]
  \nonumber \\ &&
  + \left(\frac{\alpha_s}{\pi}\right)^3
  \biggl[
    \frac{74141}{3888}
    -\frac{67127 \pi ^2}{9720}
    -\frac{41 \pi ^4}{972}
    -\frac{619}{216} \zeta_3
    +\frac{1}{4} \pi ^2 \zeta_3
    -\frac{5}{4} \zeta_5
    -\frac{8 a_4}{27}
    \nonumber \\ &&
    +\frac{640 \pi ^2 l_2}{81}
    +\frac{1}{81} \pi ^2 l_2^2
    -\frac{l_2^4}{81}
    +n_l
    \biggl(
      -\frac{5917}{11664}
      +\frac{13 \pi ^2}{324}
      +\frac{2}{27} \zeta_3
      +\Big(
        \frac{20}{27}
        -\frac{10 \pi ^2}{81}
      \Big) y
      \nonumber \\ &&
      +y^2 
      \Big(
        -\frac{4}{9}
        -\frac{\pi ^2}{36}
        -\frac{7}{12} \zeta_3
        +\frac{\pi ^2 l_2}{6}
      \Big)        
    \biggr)
    +y 
    \biggl(
      -\frac{5473}{243}
      +\frac{49738 \pi ^2}{3645}
      +\frac{979 \pi ^4}{19440}
      +\frac{839}{162} \zeta_3
      \nonumber \\ &&
      -\frac{1}{2} \pi ^2 \zeta_3
      +\frac{5}{2} \zeta_5
      +\frac{140 a_4}{27}
      -\frac{1274}{81} \pi ^2 l_2
      -\frac{35}{162} \pi ^2 l_2^2
      +\frac{35 l_2^4}{162}
    \biggr)
    \nonumber \\ &&
    +y^2 
    \biggl(
      \frac{2665}{162}
      -\frac{85549 \pi ^2}{9720}
      -\frac{979 \pi ^4}{12960}
      +\frac{473}{216} \zeta_3
      +\frac{1}{4} \pi ^2 \zeta_3
      -\frac{5}{4} \zeta_5
      -\frac{70 a_4}{9}
      +\frac{107 \pi ^2 l_2}{9}
      \nonumber \\ &&
      +\frac{35}{108} \pi ^2 l_2^2
      -\frac{35 l_2^4}{108}
    \biggr)
  \biggr]
  + \mathcal{O}(\epsilon,y^3,\alpha_s^4)\,,
\end{eqnarray}
with $y=1-x$ and for $x\to \infty$ we have
\begin{eqnarray}
  z_m^M &=&
  \left(\frac{\alpha_s}{\pi}\right)^2
  \biggl[
    -\frac{89}{432}
    +\frac{13}{18} l_z
    -\frac{1}{3} l_z^2
    +\frac{1}{225} z^2 \Bigl( 19-20 l_z \Bigr)
  \biggr]
  \nonumber \\ &&
  + \left(\frac{\alpha_s}{\pi}\right)^3
  \biggl[
    -\frac{119}{1296}
    -\frac{103 \pi ^4}{6480}
    +\frac{157}{288} \zeta_3
    +\frac{4 a_4}{9}
    -\frac{1}{54} \pi ^2 l_2^2
    +\frac{l_2^4}{54}
    \nonumber \\ &&
    +\Big(
      \frac{5755}{648}
      +\frac{2 \pi ^2}{9}
      +\frac{14}{9} \zeta_3
    \Big) l_z
    +\frac{2}{27} \pi ^2 l_2 l_z
    -\frac{287 l_z^2}{108}
    -l_z^3
    +z^2 
    \biggl(
      -\frac{3207593}{17496000}
      \nonumber \\ &&
      +\frac{667 \pi ^2}{77760}
      -\frac{29}{648} \zeta_3
      +\frac{887117 l_z}{1166400}
      -\frac{5809 l_z^2}{6480}
    \biggr)
    + n_l 
    \biggl(
      -\frac{1327}{11664}
      +\frac{2 \zeta_3}{27} 
      \nonumber \\ &&
      -\Big(
         \frac{1}{12}
        +\frac{\pi ^2}{27}
      \Big) l_z
      +\frac{2 l_z^3}{27}
      +z^2 
      \Big(
        \frac{529}{10125}
        -\frac{2 \pi ^2}{405}
        -\frac{2 l_z}{75}
        +\frac{4 l_z^2}{135}
      \Big)
    \biggr)
  \biggr]
  + \mathcal{O}(\epsilon,z^3,\alpha_s^4),
\end{eqnarray}
with $z=1/x$.

Expansion terms up to order $x^{25}$, $(1-x)^{25}$ and $(1/x)^{25}$, also to
higher order in $\epsilon$, can be found in the ancillary files~\cite{progdata}. It is
interesting to note that in the $(1-x)$ and $(1/x)$ expansion only the usual
transcendental numbers as $\zeta_n$, $\log(2)$ and $\mbox{Li}_n(1/2)$
appear. On the other hand for $x\to0$ we observe in the
$\mathcal{O}(\alpha_s^3 \epsilon)$ term Catalan's constant, see
Eq.~(\ref{eq::Catalan}).  Note that the expansions of the individual iterated
integrals show a more complicated structure.

Depending on the application it it advantageous to transform
either $m_1$ or $m_2$ or both to the $\overline{\rm MS}$  scheme.
For this purpose it is convenient to introduce the variables
\begin{eqnarray}
  x = \frac{m_2^{\rm OS}}{m_1^{\rm OS}}\,,\quad
  x_f(\mu_f) = \frac{\overline{m}_2(\mu_f)}{m_1^{\rm OS}}\,,\quad
  x_q(\mu) = \frac{m_2^{\rm OS}}{\overline{m}_1(\mu)}\,,\quad
  x_{fq}(\mu_f,\mu) = \frac{\overline{m}_2(\mu_f)}{\overline{m}_1(\mu)}\,,
\end{eqnarray}
where $\mu_f$ is the renormalization scale of the quark mass $m_2$ and $\mu$
is the common renormalization scale of $m_1$ and $\alpha_s$.  For $z_m$
explicit transformation rules to the various schemes can be found in
Ref.~\cite{Bekavac:2007tk}.
In the ancillary files to this paper we provide for 
$z_m$ and $Z_2^{\rm OS}$ different variants for the expansions in the three
limits $x\to0$, $x\to1$ and $x\to\infty$.

\begin{figure}[t]
  \includegraphics[width=\textwidth]{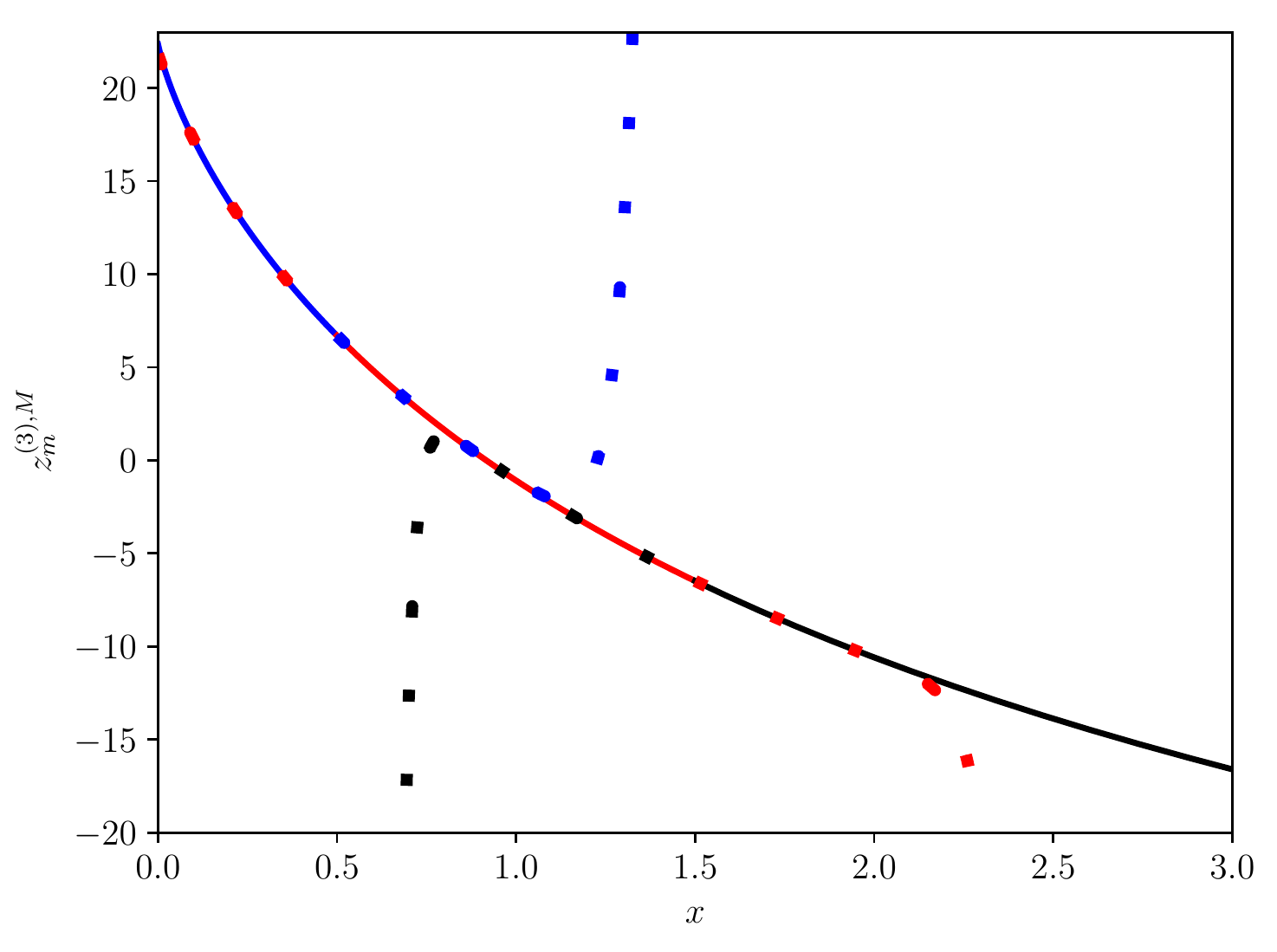}
  \caption{$z_m^{(3)}$ as a fucntion of $x$. The expansions for $x\to0$,
    $x\to1$ and $x\to\infty$ are shown as solid lines in the regions
    we the respective expansion is used for the numerical evaluation of 
    $z_m$ in the ancillary file. Outside this region dotted lines are used.}
  \label{fig:numerics}
\end{figure}

We update the {\tt Mathematica} routines provided in
Ref.~\cite{Bekavac:2007tk} for the numerical evaluation of $z_m$ and
$Z_2^{\rm OS}$. In the ancillary files to this paper one finds the functions
\verb|zmnum[x,m1,mu1,mu2[,scheme]]| and \verb|Z2OSnum[x,m1,mu1,mu2[,scheme]]| (see Appendix)
which implement
the expansion for $x\to0$, $x\to1$ and $x\to\infty$.  We switch between the
first two expansions at $x=1/2$ and between the latter two at $x=3/2$.  The
justification for this choice is illustrated in Fig.~\ref{fig:numerics}, where
we show the expansions for $z_m^{(3),M}$ for $\epsilon=0$, $n_l=3$,
$n_h=n_m=1$ and $\mu=m_1$.  In the regions where the expansions converge
($x<1/2$, $1/2<x<3/2$, $3/2<x$ for $x\to0$, $x\to1$ and $x\to\infty$, respectively) we plot
solid lines and outside these region we switch to dotted lines.  One observes
that both around $x=1/2$ and $x=3/2$ there is a large overlap among at least
two expansion, which justifies that we use the expansion results to contruct
the functions \verb|zmnum[x,m1,mu1,mu2[,scheme]]| and \verb|Z2OSnum[x,m1,mu1,mu2[,scheme]]|.
Let us also mention that we observe an agreement with the exact result to at
least 8~digits over the whole range in $x$.
Similar results are obtained for the ${\cal O}(\epsilon)$ term of $z_m$
and for $Z_2^{\rm OS}$.

To summarize, we have obtained analytic results of all 27 master integrals
which are needed to obtain the three-loop contributions for the on-shell
renormalization constants $Z_m^{\rm OS}$ and $Z_2^{\rm OS}$ with dependence 
on two different quark masses, $m_1$ and $m_2$.  Our final result includes terms
of ${\cal O}(\epsilon)$, which are relevant for a future four-loop
calculation. Furthermore, we have obtained 26 expansion terms for three cases
$m_1\gg m_2$, $m_1\approx m_2$ and $m_1\ll m_2$.



\section*{Acknowledgements}  

This research was supported by the Deutsche Forschungsgemeinschaft (DFG,
German Research Foundation) under grant 396021762 --- TRR 257 ``Particle
Physics Phenomenology after the Higgs Discovery''.


\begin{appendix}


\section{README for ancillary files}

Together with this paper we provide the following files
which contain analytic expressions for the 
various quantities in {\tt Mathematica} format:
\begin{itemize}
\item \verb|master.m| contains the 27 master 
  integrals $M_1, \ldots, M_{16}, M_{18}, \ldots, M_{28}$ as
  \verb|resM1|, \ldots, \verb|resM28|. Note that $M_{17}$ is obtained from
  Eq.~(\ref{eq::M17}).
\item \verb|zmZmZ2.m| contains exact results for the expressions \verb|zmos|,
  \verb|ZmOS| and \verb|Z2OS| in the on-shell scheme. 
  Here the variable \verb|x| corresponds to \verb|xOSOS|.
\item \verb|expansions/| is a direcory which contains the expansions for $z_m$
  and $Z_2^{\rm OS}$ in the
  three limits $x\to0$, $x\to1$ and $x\to \infty$ for various
  combinations of on-shell and $\overline{\rm MS}$ masses for $m_1$ and $m_2$.
\item \verb|zmZ2_eval.m| provides the functions \verb|zmnum[x,m1,mu1,mu2,[,scheme]]| and
  \verb|Z2OSnum[x,m1,mu1,mu2[,scheme]]| which can be used for the numerical evaluation of
   $z_m$ and $Z_2^{\rm OS}$. 
   In the case of \verb|zmnum| the option \verb|scheme| may take the values \verb|"OSOS"|, \verb|"MSOS"|,
   \verb|"OSMS"| and \verb|"MSMS"|, where the first (last) two
   letters refer to the scheme of $m_2$ ($m_1$). In the case of \verb|Z2OSnum|
   the values \verb|"OSOS"| and \verb|"MSOS"| are allowed.
   Depending on the value of $x$ and the specified
   scheme the corresponding results from \verb|expansions/| are loaded.
\item \verb|toGINAC.m| provides rules which maps the iterated integrals
  \verb|GL[{...}, x]| and HPLs \verb|H[..., x]| to Goncharov polylogarithms 
  which allows for a numerical evaluation with {\tt GiNaC}~\cite{Bauer:2000cp}.
\end{itemize}
For the meaning of the symbols we refer to Tab.~\ref{tab::symbols}.
The exact expressions in \verb|zmZmZ2.m| contain in addition
the iterated integrals
\verb|GL[{...}, x]| $ = I(\{...\}),x)$ (cf. Eq.~(\ref{eq::it_int}))
and the HPLs (\verb|H[..., x]|), both
in the notation of {\tt HamonicSums}~\cite{HarmonicSums}.

\begin{table}[t]
  \begin{center}
  \begin{tabular}{c|c| c|c| c|c| c|c}
    api &  lmm1 &  cf & ca & tr & nl & nm & nh 
    \\ \hline
    $\alpha_s(\mu)/\pi$ & $\log(\mu^2/(m_1^{\rm OS})^2)$ & 
    $C_F$ & $C_A$ & $T_F$ & $n_l$ & $n_m$ & $n_h$
    \\ \hline \hline
    x & xOSOS & xMSOS & xOSMS & xMSMS & mu1 & mu2
    \\ \hline
    $x$ & $x$ & $x_f(\mu_f)$ & $x_q(\mu_f)$ & $x_{fq}(\mu_f,\mu)$ & $\mu$ & $\mu_f$
    \\ \hline \hline
    m1OS & m1MS & m2OS & m2MS &  &  & 
    \\ \hline
    $m_1^{\rm OS}$ & $\overline{m}_1(\mu)$ & $m_2^{\rm OS}$ & $\overline{m}_2(\mu_f)$ & & & 
  \end{tabular}
  \caption{\label{tab::symbols}Meaning of the symbols used in the {\tt Mathematica} expressions.}
  \end{center}
\end{table}


\end{appendix}




\begin{thebibliography}{99}

%
%

\bibitem{Bekavac:2007tk}
S.~Bekavac, A.~Grozin, D.~Seidel and M.~Steinhauser,
JHEP \textbf{10} (2007), 006
%
[arXiv:0708.1729 [hep-ph]].

\bibitem{Tarrach:1980up}
R.~Tarrach,
Nucl. Phys. B \textbf{183} (1981), 384-396.

\bibitem{Gray:1990yh}
N.~Gray, D.~J.~Broadhurst, W.~Grafe and K.~Schilcher,
Z. Phys. C \textbf{48} (1990), 673-680.

\bibitem{Broadhurst:1991fy}
D.~J.~Broadhurst, N.~Gray and K.~Schilcher,
Z. Phys. C \textbf{52} (1991), 111-122.

\bibitem{Chetyrkin:1999ys}
K.~G.~Chetyrkin and M.~Steinhauser,
Phys. Rev. Lett. \textbf{83} (1999), 4001-4004
[arXiv:hep-ph/9907509 [hep-ph]].

\bibitem{Chetyrkin:1999qi}
K.~G.~Chetyrkin and M.~Steinhauser,
Nucl. Phys. B \textbf{573} (2000), 617-651
[arXiv:hep-ph/9911434 [hep-ph]].

\bibitem{Melnikov:2000qh}
K.~Melnikov and T.~v.~Ritbergen,
Phys. Lett. B \textbf{482} (2000), 99-108
[arXiv:hep-ph/9912391 [hep-ph]].

\bibitem{Melnikov:2000zc}
K.~Melnikov and T.~van Ritbergen,
Nucl. Phys. B \textbf{591} (2000), 515-546
[arXiv:hep-ph/0005131 [hep-ph]].

\bibitem{Marquard:2007uj}
P.~Marquard, L.~Mihaila, J.~H.~Piclum and M.~Steinhauser,
Nucl. Phys. B \textbf{773} (2007), 1-18
[arXiv:hep-ph/0702185 [hep-ph]].

\bibitem{Marquard:2015qpa}
P.~Marquard, A.~V.~Smirnov, V.~A.~Smirnov and M.~Steinhauser,
Phys. Rev. Lett. \textbf{114} (2015) no.14, 142002
[arXiv:1502.01030 [hep-ph]].

\bibitem{Marquard:2016dcn}
P.~Marquard, A.~V.~Smirnov, V.~A.~Smirnov, M.~Steinhauser and D.~Wellmann,
Phys. Rev. D \textbf{94} (2016) no.7, 074025
[arXiv:1606.06754 [hep-ph]].

\bibitem{Marquard:2018rwx}
P.~Marquard, A.~V.~Smirnov, V.~A.~Smirnov and M.~Steinhauser,
Phys. Rev. D \textbf{97} (2018) no.5, 054032
[arXiv:1801.08292 [hep-ph]].

\bibitem{Laporta:2020fog}
S.~Laporta,
Phys. Lett. B \textbf{802} (2020), 135264
%
[arXiv:2001.02739 [hep-ph]].

\bibitem{progdata}
\verb|https://www.ttp.kit.edu/preprints/2020/ttp20-028/|.

\bibitem{Bekavac:2009gz}
S.~Bekavac, A.~G.~Grozin, D.~Seidel and V.~A.~Smirnov,
Nucl. Phys. B \textbf{819} (2009), 183-200
%
[arXiv:0903.4760 [hep-ph]].

\bibitem{Lee:2012cn}
R.~N.~Lee,
[arXiv:1212.2685 [hep-ph]].

\bibitem{Lee:2013mka}
R.~N.~Lee,
J. Phys. Conf. Ser. \textbf{523} (2014), 012059
[arXiv:1310.1145 [hep-ph]].

\bibitem{Ablinger:2018zwz}
J.~Ablinger, J.~Bl\"umlein, P.~Marquard, N.~Rana and C.~Schneider,
Nucl. Phys. B \textbf{939} (2019), 253-291
[arXiv:1810.12261 [hep-ph]].

\bibitem{ORESYS}
S.~Gerhold, {\it Uncoupling systems of linear {O}re operator equations},
Master's thesis, RISC, J.~Kepler University, Linz, 2002. 
%
%

\bibitem{Schneider:2007}
C.~Schneider, {S\'em.~Lothar. Combin.\/} {\bf 56} (2007) 1,
 article B56b;
C.~Schneider, in:~{{Computer Algebra in Quantum Field Theory: Integration,
  Summation and Special Functions}\/} Texts and Monographs in Symbolic
  Computation eds. C.~Schneider and J.~Bl\"umlein  (Springer, Wien, 2013) 325
  arXiv:1304.4134 [cs.SC].

\bibitem{HarmonicSums}
J.~Vermaseren,
Int. J. Mod. Phys. A \textbf{14} (1999), 2037-2076
%
[arXiv:hep-ph/9806280 [hep-ph]];
%
%
%
%
J.~Bl\"umlein,
Comput. Phys. Commun. \textbf{180} (2009), 2218-2249
%
[arXiv:0901.3106 [hep-ph]];
  J.~Ablinger,
  Diploma Thesis, J. Kepler University Linz, 2009,
  arXiv:1011.1176 [math-ph];
  J.~Ablinger, J.~Bl\"umlein and C.~Schneider,
  J.\ Math.\ Phys.\  {\bf 52} (2011) 102301
  [arXiv:1105.6063 [math-ph]];
J.~Ablinger, J.~Bl\"umlein and C.~Schneider,
J. Math. Phys. \textbf{54} (2013), 082301
%
[arXiv:1302.0378 [math-ph]];
  J.~Ablinger,
  Ph.D. Thesis, J. Kepler University Linz, 2012,
  arXiv:1305.0687 [math-ph];
J.~Ablinger, J.~Bl\"umlein and C.~Schneider,
J. Phys. Conf. Ser. \textbf{523} (2014), 012060
%
[arXiv:1310.5645 [math-ph]];
J.~Ablinger, J.~Bl\"umlein, C.~Raab and C.~Schneider,
J. Math. Phys. \textbf{55} (2014), 112301
%
[arXiv:1407.1822 [hep-th]];
%
J.~Ablinger,
PoS \textbf{LL2014} (2014), 019
%
[arXiv:1407.6180 [cs.SC]];
%
J.~Ablinger,
[arXiv:1606.02845 [cs.SC]];
%
J.~Ablinger,
PoS \textbf{RADCOR2017} (2017), 069
[arXiv:1801.01039 [cs.SC]];
%
J.~Ablinger,
PoS \textbf{LL2018} (2018), 063;
%
%
J.~Ablinger,
[arXiv:1902.11001 [math.CO]].
%

\bibitem{Kovacic}  
J.J.~Kovacic,
J.~Symbolic Comput. {\bf 2}, 1986.
%
%

\bibitem{Lee:2010ik}
R.~N.~Lee and V.~A.~Smirnov,
JHEP \textbf{02} (2011), 102
[arXiv:1010.1334 [hep-ph]].
%

\bibitem{Remiddi:1999ew}
E.~Remiddi and J.~A.~M.~Vermaseren,
Int. J. Mod. Phys. A \textbf{15} (2000), 725-754
[arXiv:hep-ph/9905237 [hep-ph]].

\bibitem{Ablinger:2011te}
J.~Ablinger, J.~Bl\"umlein and C.~Schneider,
J. Math. Phys. \textbf{52} (2011), 102301
[arXiv:1105.6063 [math-ph]].

\bibitem{Goncharov:1998kja}
  A.~B.~Goncharov,
  Math.\ Res.\ Lett.\  {\bf 5} (1998) 497
  [arXiv:1105.2076 [math.AG]].

\bibitem{Bauer:2000cp}
C.~W.~Bauer, A.~Frink and R.~Kreckel,
J. Symb. Comput. \textbf{33} (2002), 1-12
%
[arXiv:cs/0004015 [cs.SC]].


\end{thebibliography}
\end{document}